\documentclass[preprint]{aastex}
\shorttitle{Revisiting NSVS14256825}
\shortauthors{Hinse et al.}

\usepackage{pslatex,color}
\usepackage{amsmath,amsfonts,amssymb,amsxtra,eufrak,url}

\begin{document}

\date{MNRAS, Accepted 2013 November 7.  Received 2013 November 5; in original form 2013 October 11}

\title {Revisiting the proposed circumbinary multi-planet system NSVS14256825}

\author{Tobias Cornelius Hinse \& Jae Woo Lee}
\email{tchinse@gmail.com}

\affil{Korea Astronomy and Space Science Institute, Daejeon 305-348, Republic of Korea}

\author{Krzysztof Go{\'z}dziewski}
\affil{Nicolaus Copernicus University, Torun Centre for Astronomy, PL-87-100 Torun, Poland}
	
\author{Jonathan Horner \& Robert A. Wittenmyer}
\affil{Department of Astrophysics and Optics, School of Physics, University of New South Wales, Sydney, 2052 Australia}

\begin{abstract}
In this work we carry out an analysis of the observed times of primary and secondary eclipses of the post-common envelope binary NSVS14256825. Recently, \cite{Almeida2013} proposed that two circumbinary companions orbit this short-period eclipsing binary, in order to explain observed variations in the timing of mutual eclipses between the two binary components. Using a standard weighted least-squares minimisation technique, we have extensively explored 
the topology of $\chi^2$ parameter space of a single planet model. We find the data set to be insufficient to reliably constrain a one-companion model. 
Various models, each with similar statistical significance, result in substantially different orbital architectures for the additional companion. No evidence is seen for a second companion of planetary nature. We suspect insufficient coverage (baseline) of timing data causing the best-fit parameters to be unconstrained.
\end{abstract}

{}

\section{Introduction}

The discovery of planets within binary star systems has recently sparked an increased interest in their formation, occurrence frequency and dynamical evolution \citep{Zwart2013}. Several techniques exist to detect additional bodies accompanying binary stars. In addition to the traditional radial velocity technique, \cite{Han2008} outlines the possibility to infer such planets from microlensing observations. Recently, transiting circumbinary planets have been detected using the Kepler space telescope \citep{Doyle2011, Welsh2012, Orosz2012a, Orosz2012b, Schwamb2013, Kostov2013}. Furthermore, companions can 
be detected from pulsar timing measurements \citep{WolszczanFrail1992}. The formation and dynamical evolution of planets around binary star systems have been the subject of recent theoretical studies \citep{QuintanaLissauer2006, HaghighipourRaymond2007, Marzari2009, Shi2012}.

Utilising ground-based observations, a number of multi-planet systems around short-period eclipsing binary stars have been proposed in recent years \citep{Lee2009,Beuermann2010,Potter2011,Qian2011}. From measuring the times at minimum light (either primary and/or secondary eclipse) one can use the light-travel time (LTT) effect to detect additional companions by measuring periodic changes in the binary period \citep{Irwin1952,Hinse2012a,Horner2012a}. In contrast to other detection methods (radial velocity, microlensing and transit) the LTT technique is sensitive to massive companions on a long-period orbit: the semi-amplitude $K$ of the LTT signal scales with the companions mass and period as $K \sim M_3$ and $K \sim P_{3}^{2/3}$, respectively. In addition, low-mass binary components will favour the detection of low-mass companions on short-period orbits \citep{Pribulla2012}.

From ground-based photometric observations, the first two-planet circumbinary system (HW Virginis, a.k.a HW Vir) was proposed by \cite{Lee2009}. Additional multi-body systems of planetary nature were subsequently proposed by \cite{Beuermann2010, Marsh2013} (NN Serpentis, a.k.a NN Ser), \cite{Potter2011} (UZ Fornacis, a.k.a. UZ For) and \cite{Qian2011} (HU Aquarii, a.k.a. HU Aqr). Recently, \cite{Lee2012} proposed a quadruple system with two circumbinary sub-stellar companions orbiting the Algol-type binary SZ Hercules 
(a.k.a. SZ Her).

For a secure detection of a multi-planet circumbinary system, at least two criteria need to be satisfied. First, any period variation, due to additional companions, must be recurring and periodic in time. The data should extend over at least two complete cycles of the longest period. Second, the proposed system 
should be dynamically stable on time scales comparable to the age of the 
binary components. \cite{Horner2011} first studied the dynamical stability of the two planets in HU Aqr. Their study allowed them to conclude that the system is highly unstable with disruption times of a few hundred years. Subsequent studies of the same system were carried out by \cite{Hinse2012a}, \cite{Wittenmyer2012} and \cite{Gozdziewski2012}. The overall conclusion of these studies is that the planets, as proposed in the discovery work, are simply not feasible. More observational data is necessary before any further constraints can be imposed on the orbital parameters of any companions in the HU Aqr system. 

The proposed planets orbiting the close binary system HW Vir \citep{Lee2009} is another case where the proposed planets do not stand up to dynamical scrutiny \citep{Horner2012b}. In that case, the dynamical character of the HW Vir system was studied, and the planets proposed were found to follow highly unstable orbits most likely due to their crossing orbit architecture and relatively high masses. However, \cite{Beuermann2012b} presented new timing measurements of HW Vir allowing them to conclude stable orbits under the assumption of fixing some of the orbital elements in their least-square analysis.

The NN Ser system was also recently studied by \cite{Horner2012a}. These authors found stable orbits for the \cite{Beuermann2010} solutions, if the planets are locked in a mean-motion resonant (MMR) configuration. However, an in-depth remodeling of the timing data renders the system unstable when all parameters are allowed to vary freely. Very recently, \cite{Beuermann2013} published additional timing data of NN Ser. Their re-analysis allowed them to conclude 
the existence of two companions orbiting the binary pair involved in a 2:1 MMR.

Unstable orbits in proposed multi-body circumbinary systems have not only been found among companions of planetary nature. The SZ Her system with two sub-stellar mass companions was recently investigated within a dynamical analysis \citep{Hinse2012b}. Here, the authors also found that the proposed companions followed highly unstable orbits.

In a recent work, \cite{Almeida2013} interpreted observed eclipse timing variations of the post-common envelope binary NSVS14256825 as being the result of a pair of light-travel time effect introduced by two unseen circumbinary companions. The proposed companions are of planetary nature, with orbital periods $\simeq 3.5$ and $\simeq 6.7$ years, and masses of $3~M_{jup}$ and $8~M_{jup}$, respectively. Once again, however, a recent study \citep{Wittenmyer2013} reveals that the proposed planetary system would be dynamically unstable on very short timescales - with most plausible orbital architectures being unstable on timescales of just a few hundred years, and only a small fraction of systems surviving on timescales of $10^5$ years.

The aim of this paper is as follows. In section 2 we present the available timing data of NSVS14256825, which forms the basis of our analysis. In particular we augment the timing measurements presented in \cite{Almeida2013} with three additional data points presented in \cite{Beuermann2012a}. We also introduce the light-travel time model using Jacobian coordinates and outline the derivation of the minimum mass and projected semi-major axis for a single circumbinary companion along with a short description of our least-squares minimisation methodology. In section 3, we carry out a data analysis and 
perform a period analysis based on Fourier techniques and present our results describing the main properties of our best-fit solutions. In particular, we present results from finding a best-fit linear, best-fit quadratic and best-fit one-companion model. Finally, we summarise our results and discuss our conclusion in section 4.

\section{Data acquisition and Jacobian light-travel time model}
\label{dataacquisition}
As the basis of this work we consider the same timing data set as published in \cite{Almeida2013}. However, we noticed that three timing measurements published in \cite{Beuermann2012a} were not included in \cite{Almeida2013}. We have therefore carried out two independent analysis based on the following data sets. \emph{Dataset I}: Data as presented in Table 3 in \cite{Almeida2013}. This data set spans the period from June 22, 2007 to August 13, 2012, corresponding to an observing baseline of around 5 years.
\emph{Dataset II}: Data as presented in Table 3 in \cite{Almeida2013} \emph{plus
three data points} (primary eclipse) from \cite{Beuermann2012a}. The additional points are as follows. BJD $2,451,339.803273 \pm 0.000429$ days, BJD 
$2,452,906.673899 \pm 0.000541$ days and BJD $2,453,619.579776 \pm 0.000537$ days. The second data set spans the period from June 10, 1999 to August 13, 2012, corresponding to an observing baseline of around 13 years (i.e., doubling the time window).

The aim of considering the second data set (Dataset II) is to investigate the effect of the additional timing data on the overall best-fit solution and compare the results obtained from considering the first data set (Dataset I) since it covers a longer observing baseline.

The time stamps in \cite{Beuermann2012a} are stated using the terrestrial time (TT) standard while the times in \cite{Almeida2013} states timing measurements in the barycentric dynamical time (TDB) standard. However, the difference between these time standards (TT vs TDB) introduces timing differences on a milli-second (approx. 0.002s) level due to relativistic effects \citep{Eastman2010}. In light of the quoted measurement uncertainties (from the literature) of the eclipse timings in the two data sets, the two time stamps (TT and TDB) can be combined and no further transformation of one time standard to the other is necessary.

Considering the binary as an isolated two-body system and in the absence of mechanisms that cause period variations, the linear ephemeris of future (or past) eclipses $T_{ecl}$ is given by \citep{Hilditch2001}

\begin{equation}
T_{ecl}(E) = T_{0} + E \times P_{0},
\label{linearephemerisequation}
\end{equation}
\noindent
where $E$ denotes the cycle number, $T_0$ is the reference epoch and $P_0$ is the nominal binary period. Additional effects that cause variations of the binary period would be observed as a systematic residual about this best-fit line.

We use the formulation of the light-travel time effect based on Jacobian coordinates \citep{Gozdziewski2012}. In the general case a circumbinary 
$N$-body system is a hierarchical system and employing Jacobian elements therefore seems natural. This is particularly true for the case of a single companion (the first object in a hierarchical multi-body ensemble), where the Jacobian coordinate is equivalent to astrocentric coordinates and readily returns the geometric osculating orbital elements of the companion relative to the binary. Here we assume the binary to be a single massive object with mass equivalent to the sum of the two component masses. For a single circumbinary companion the LTT signal can be expressed as \citep{Gozdziewski2012}

\begin{equation}
\tau(t) = -\frac{\zeta_1}{c},
\label{equation2}
\end{equation}
\noindent
where $c$ is the speed of light and $\zeta_{1}$ is given as
\begin{equation}
\zeta_{1}(t) = K_{1}\Bigg[ \sin\omega_{1} (\cos E_{1}(t) - e_{1}) + \cos\omega_{1} \sqrt{1-e_{1}^2} \sin E_{1}(t)\Bigg],
\end{equation}
\noindent
Here $e_{1}$ denotes the orbital eccentricity and $\omega_1$ measures the argument of pericentre of the companion relative to the combined binary representing the dynamical centre. The eccentric anomaly is given as $E_{1}$. Following \cite{Gozdziewski2012} the semi-amplitude of the LTT signal is given as
\begin{equation}
K_{1} = \Bigg( \frac{1}{c}\Bigg)~\frac{m_1}{m_{*} + m_1}~a_{1}\sin I_{1},
\end{equation}
\noindent
with $c$ measuring the speed of light, $a_{1}$ the semi-major axis, $I_{1}$ the inclination of the orbit relative to the skyplane. The quantities $m_{*}$ and $m_{1}$ denote the masses of the combined binary and companion, respectively. 

In summary, the set $(K_{1},P_{1},e_{1},\omega_{1}, T_{1})$ represent the five free osculating orbital parameters for the companion with $P_{1}$ and $T_{1}$ denoting the orbital period and time of pericentre passage, respectively. These latter two quantities are introduced implicitly via Kepler's equation and the eccentric anomaly \citep{Gozdziewski2012,Hinse2012a}

\subsection{Deriving minimum mass and projected semi-major axis}

Once a weighted least-squares best-fit model has been found the minimum mass 
of the companion is obtained from solving the following transcendental function

\begin{equation}
f(m_{1}) = \gamma_{1}(m_{1}+m_{*})^2 - m_{1}^3 = 0,
\end{equation}
\noindent
where
\begin{equation}
\gamma_{1} = \Bigg( \frac{c^3}{k^2}\Bigg) \Bigg( \frac{4\pi^2}{P_1^2}\Bigg)K_{1}^3.
\end{equation}
\noindent
The projected minimum (with $\sin I_{1} = 1$) semi-major axis ($a_1$) is 
then found from Kepler's third law
\begin{equation}
\frac{P_{1}^2}{a_{1}^3} = \frac{4\pi^2}{\mu_1},
\end{equation}
\noindent
where the gravitational parameter is given by $\mu_1 = k^2 (m_{*}+m_{1})$ with $k$ denoting Gauss' gravitational constant. The combined mass of the two binary components is assumed to be $m_{*} = 0.528 M_{\odot}$ \citep{Almeida2013}.

Considering only the case of a single circumbinary companion, the timings of minimum light for primary eclipses is given as
\begin{equation}
T_{ecl}(E) = T_{0} + E \times P_{0} + \tau(K_{1}, P_{1}, e_{1}, \omega_{1}, T_{1}).
\end{equation}
\noindent
We therefore have a total of seven model parameters describing the light-travel time effect caused by a single circumbinary companion. For a description of two companions we refer to \cite{Gozdziewski2012}. The LTT signal is a one-dimensional problem similar to the radial velocity 
technique. We therefore only derive the minimum mass and minimum (projected) semi-major axis of the companion. For simplicity, we henceforth write $m_{1}$ and $a_{1}$ for the minimum masses and minimum semi-major axis of the companion\footnote{Technically, the values obtained represent the minimum possible values of $m_{1}\sin I_{1}$ and $a_{1}\sin I_{1}$ - but in standard papers dealing with eclipse timing or radial velocity studies authors use the shortened versions, for brevity.}.

It is worth pointing out that no gravitational interactions have been taken into account in the above formulation of the LTT signal. Only Keplerian motion is considered. It is possible to include additional effects (such as mutual gravitational interactions) that can cause period variations and we refer to \cite{Gozdziewski2012} for more details.

Finally, we stress that the case of a {\it single} companion the Jacobian-based description of the one-companion LTT effect is equivalent to the formulation given in \cite{Irwin1952, Irwin1959}. Hence, the $P_{1}, e_{1}, \omega_{1}, T_{1}$ should be identical to those parameters obtained using the \cite{Irwin1952} LTT model along with the derived minimum mass. The only parameter which is different is the semi-major axis of the binary due to the different reference systems used and we refer the reader to \cite{Gozdziewski2012} for details. For consistency, we tested our results for the presently (Jacobian) derived LTT formulation using the procedure detailed in \cite{Irwin1952}, and obtained identical results. However, one complication could arise in the argument of pericentre which can differ depending on the defined direction of the line-of-sight axis. Either this axis can point towards or away from the observer. The difference will affect the argument of pericentre and can be rectified using the relation $\omega_1 = \omega^{'} + \pi$, where $\omega^{'}$ is the argument of pericentre defined in a reference system with opposite line-of-sight direction compared to the formulation outlined in \cite{Gozdziewski2012}. Hence the difference is only a matter of convention and does not affect the quantitative results obtained from the two formulations.

\subsection{Weighted least-squares fitting}

We have implemented the Jacobian-based Kepler-kinematic LTT model in IDL\footnote{http://www.exelisvis.com/ProductsServices/IDL.aspx}. The Levenberg-Marquardt (LM) least-square minimisation algorithm was used to find a best-fit model and is available via the \texttt{MPFIT} routine \citep{Markwardt2009}. We quantify the goodness of fit statistic as

\begin{equation}
\chi^2 = \sum_{i=1}^{N}\Bigg( \frac{O_{i}-C_{i}}{\sigma_{i}}\Bigg)^2,
\label{chisquare}
\end{equation}
\noindent
where $N$ is the number of data points, $O_{i}-C_{i}$ measures the vertical difference between the observed data and the computed model at the $i$th cycle, and $\sigma_{i}$ measures the 1-sigma timing uncertainty (usually obtained formally). However, in this work we will quote the \emph{reduced} 
$\chi^2$ defined as $\chi^2_{\nu} = \chi^2/\nu$ with $\nu = N-n$ denoting the degree of freedom. The \texttt{MPFIT} routine attempts to minimise $\chi^2$ iteratively using $n$ free parameters.

In the search for a global minimum $\chi^2_{\nu,0}$ of the underlying 
$\chi^2$ space we utilise a Monte Carlo approach by generating a large number $(5 \times 10^5)$ of random initial guesses. Two approaches can be used to explore the $\chi^2$ space for a global minimum. The first involves generating random initial guesses in a relatively narrow region of a given parameter and may be applied when information about the periodicity and 
amplitude of the LTT signal is inferred from other means (e.g. Fourier analysis). For example, if a Fourier analysis reveals a given frequency within the data one can then generate random initial guesses from a normal distribution centred at that period with some (more or less narrow) standard deviation for the variance. In the second approach, random initial guesses are generated from a uniform distribution defined over a broad interval for a given parameter. However, in both approaches we randomly choose the eccentricity from a uniform distribution within $e_1 \in [0,0.99]$ with the argument of pericentre chosen from $\omega_1 \in [-\pi,\pi]$. In all our searches we recorded the initial guess and final parameters along with the goodness-of-fit value, the corresponding root-mean-square (RMS) statistic and formal 1-sigma uncertainties. A single LM iteration sequence is terminated following default values of accuracy parameters within \texttt{MPFIT} or after a maximum of 3000 iterations (rarely encountered with the average number of iterations required being just 11).

\section{Data analysis \& results}

\subsection{Period analysis and linear ephemeris}

\begin{figure*}
\centering
\includegraphics[angle=0,scale=1.0]{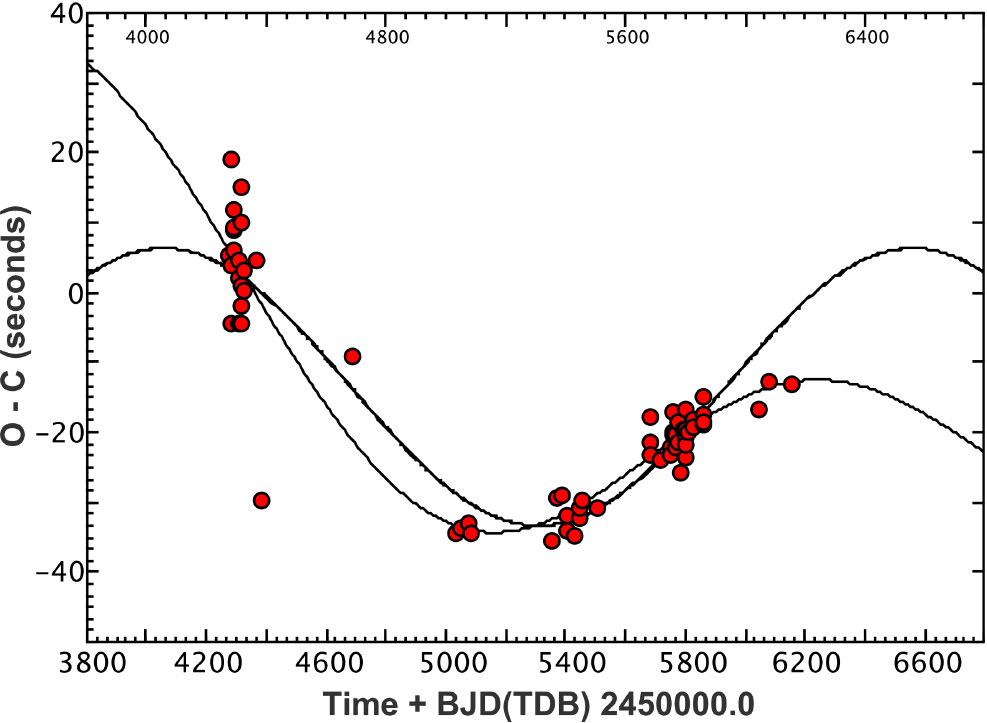}
\caption{Graphical result from PERIOD04 analysis performed on Dataset I showing two graphs with one or two Fourier components. The general form follows $O-C = \sum A_{i} \sin(2\pi (\omega_i t + \phi_i))$. The first component has $A_1 = 19.85$ seconds, $\omega_1 = 3.99\times 10^{-4}$ cycles/days (corresponding to a period of 6.9 years) and $\phi_1 = 0.63$ radians. The second component has $A_2 = 33.37$ seconds, $\omega_2 = 1.33\times 10^{-4}$ cycles/day (period of 20.6 years) and $\phi_2 = 0.93$ radians. {\it See electronic version for colors}.}
\label{periodstudy}
\end{figure*}

\begin{figure*}
\centering
\includegraphics[angle=0,scale=1.0]{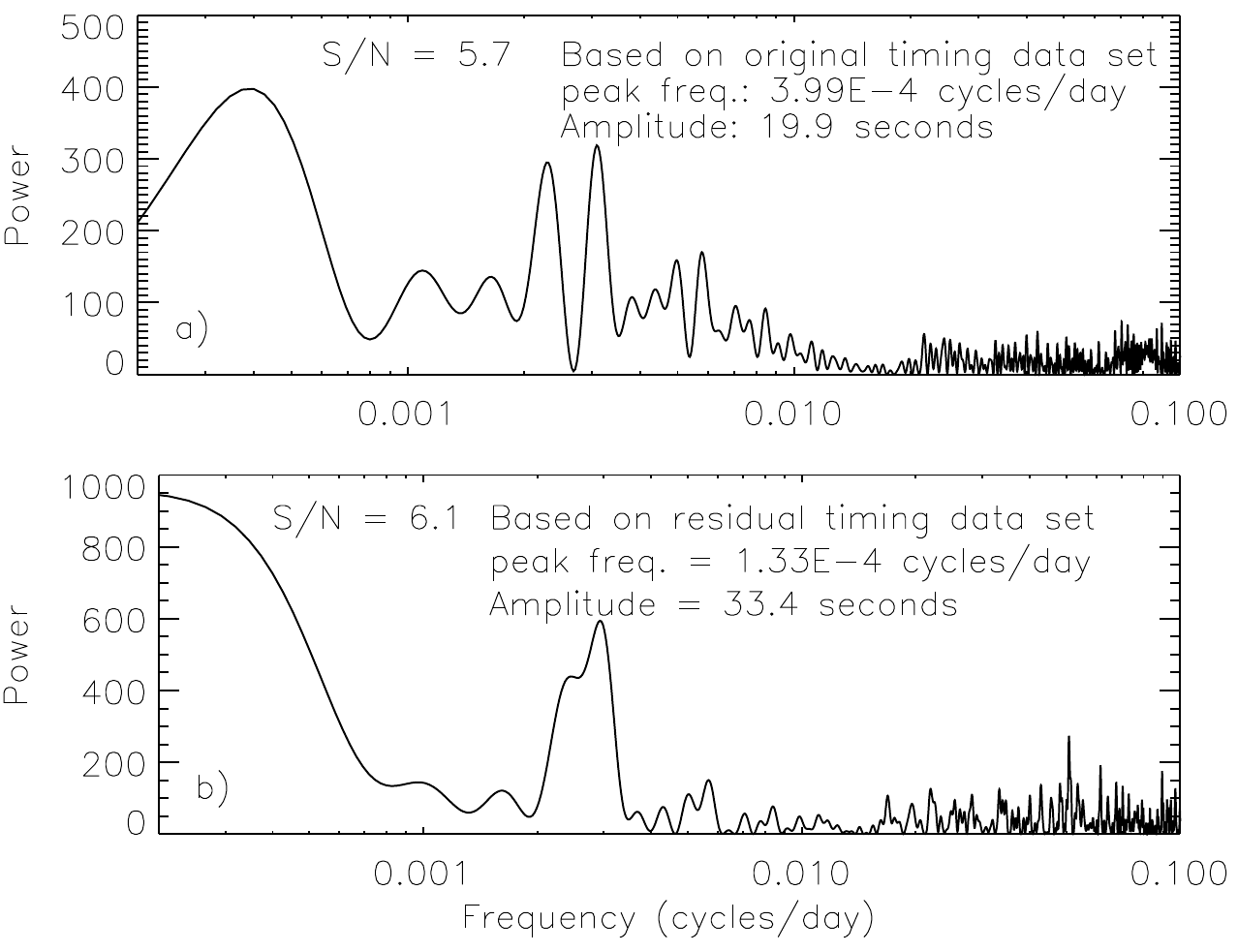}
\caption{Power spectrum of the NSVS14256825 timing data with the linear part (from a linear regression) subtracted. Additional peaks correspond to $\simeq 1$ year alias frequencies originating from the annual observing cycle.} 
\label{powerspec}
\end{figure*}

As a starting point for our analysis, we first determined the parameters of the linear ephemeris ($T_{0}, P_{0}$) by calculating a linear least-squares regression line to the same data (Dataset I) as considered by \cite{Almeida2013}. A best fit line resulted in a $\chi^2_{65} \simeq 13$ with $n=2$ free parameters and $N=67$ data points. The corresponding $\chi^2$ value was found to be 853 and the (rounded) linear ephemeris was determined to be

\begin{eqnarray}
T_{ecl}^{I}&=& T_{0} + E \times P_{0} \\
       &=& \textnormal{BJD}~2455408.744502 \pm 3 \times 10^{-6} + E \times 0.1103741881 \pm 8 \times 10^{-10}~\textnormal{days} \\
       &=& \textnormal{BJD}~2455408.744502^{505}_{499} + E \times 0.1103741881^{89}_{73}~\textnormal{days}
\end{eqnarray}
\noindent
For Dataset II we obtained the slightly different ephemeris, with little improvement in the precision of the binary period
\begin{eqnarray}
T_{ecl}^{II}&=& T_{0} + E \times P_{0} \\
           &=& \textnormal{BJD}~2455408.744504 \pm 3 \times 10^{-6} + E \times 0.1103741759 \pm 8 \times 10^{-10}~\textnormal{days} \\
           &=& \textnormal{BJD}~2455408.744504^{507}_{501} + E \times 0.1103741759^{67}_{51}~\textnormal{days}
\end{eqnarray}
\noindent
We applied the \texttt{PERIOD04}\footnote{http://www.univie.ac.at/tops/Period04/} \citep{LenzBreger2005} Lomb-Scargle algorithm on the residual data 
(Fig.~\ref{periodstudy}) obtained from subtracting the best-fit line, and compared two fits to the residual data. The first had a single Fourier component, whilst the second had two Fourier components as shown in Fig.~\ref{periodstudy}. The two-component fit was found to provide a better description of the data. We show the corresponding power spectra in Fig.~\ref{powerspec}, and find the 6.9 year period to be in agreement with the period found by \cite{Almeida2013}. However, the algorithm was unable to detect the 3.5 year cycle (inner proposed planet) as determined in \cite{Almeida2013}. Instead, we found a 20.6 year cycle with a detection six times above the noise level.

\subsection{Quadratic ephemeris model - Dataset I}
\label{quadraticephemerissection}

\begin{figure*}
\includegraphics[angle=0,scale=0.45]{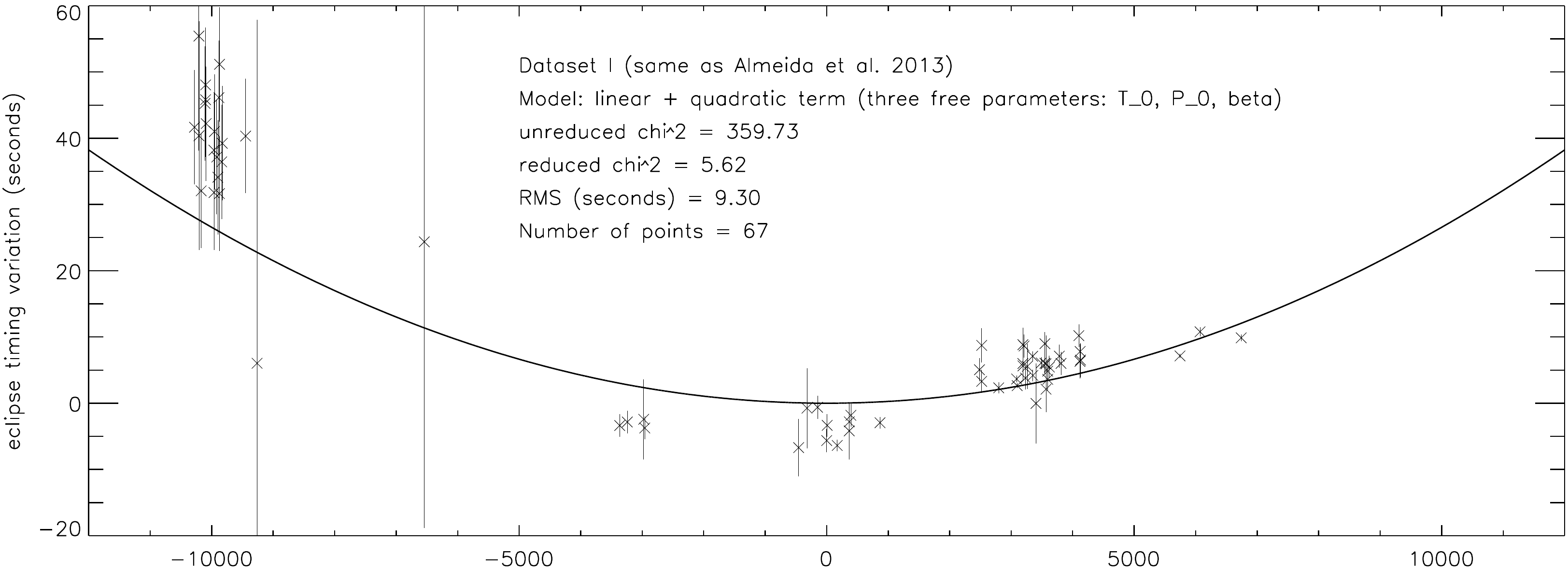}
\includegraphics[angle=0,scale=0.45]{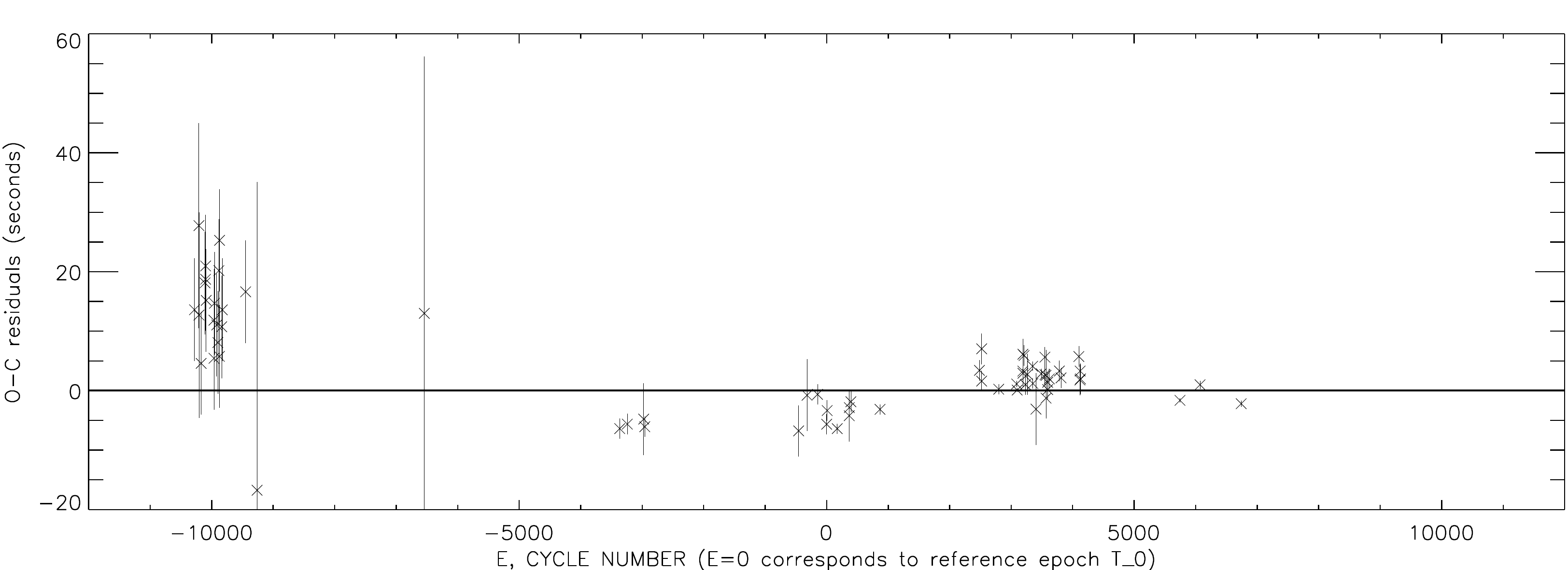}
\caption{Quadratic ephemeris model to Dataset I. Best-fit parameters ($T_{0}, P_{0}, \beta$) along with their formal uncertainties are listed in section \ref{quadraticephemerissection}. The loci of points at $E = -10,000$ appear to be systematically off-set by +15 seconds from the expected parabola. The root-mean-square scatter around the parabola is around 9 seconds.} 
\label{SecularBestFitModel}
\end{figure*}

\begin{figure*}
\includegraphics[angle=0,scale=1.0]{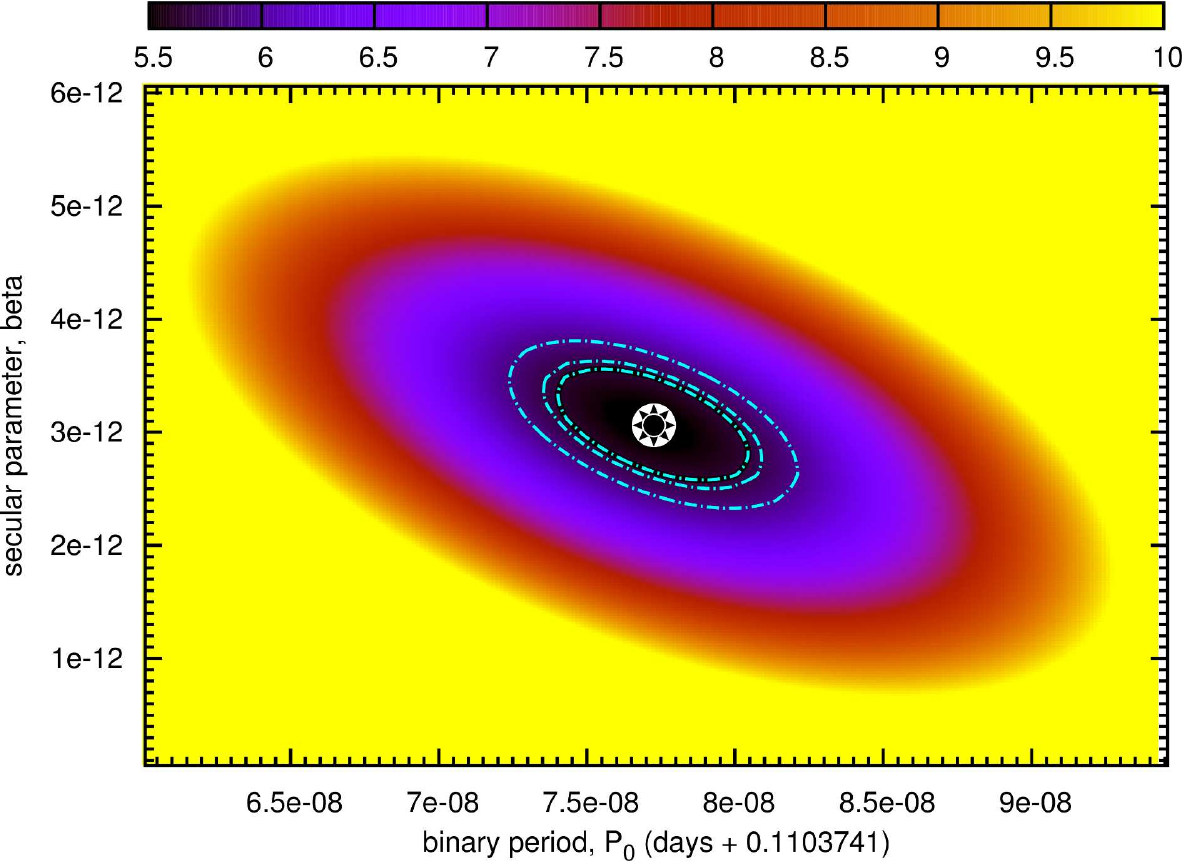}
\caption{Color-coded $\chi^2_{\nu}$ scans for the quadratic ephemeris model (Fig.~\ref{SecularBestFitModel}). The $\beta$ parameter denotes the period damping factor. Remaining parameters were allowed to vary freely. The best-fit solution is shown by a star-like symbol. Contour curves (from inner to outer) show the $1-, 2-, 3-\sigma$ confidence levels around the best-fitting model (symbol). See text for more details. {\it See electronic version for colors}.}
\label{2DMapSecularBestFitModel}
\end{figure*}

In some cases a change of the binary period can be caused by non-gravitational
interaction between the two components of a short-period eclipsing binary. Often the period change is described by a quadratic ephemeris (linear plus secular) with the times of primary eclipses given by \cite{Hilditch2001}
\begin{equation}
T_{ecl}(E) = T_{0} + P_{0} \times E + \beta \times E^{2},
\end{equation}
\noindent
where $\beta$ is a period damping factor \citep{Gozdziewski2012} which can account for mass-transfer, magnetic braking, gravitational radiation and/or the influence of a distant companion on a long-period orbit. Following \cite{Brinkworth2006} the rate of period change, in the case of mass-transfer, is then given by
\begin{equation}
\dot{P} = \frac{2\beta}{P},
\end{equation}
\noindent 
with $P$ denoting the currently measured binary period. In Fig.~\ref{SecularBestFitModel} we show the best-fit quadratic ephemeris 
given as
\begin{eqnarray}
T_{ecl}(E) = (\textnormal{BJD}~2,455,408.744485 \pm 3.4 \times 10^{-6}) &+& (0.1103741772 \pm 8.9 \times 10^{-10}) \times E \\ &+& (3.1 \times 10^{-12} \pm 1.4 \times 10^{-13}) \times E^{2}
\end{eqnarray}
\noindent
with unreduced $\chi^2 = 360$ for (67-3) degrees of freedom. In Fig.~\ref{2DMapSecularBestFitModel} we show the location of the best-fit surrounded by the $1\sigma$ (68.3\%, $\Delta\chi^2 = 2.3$), $2\sigma$ (95.4\%, $\Delta\chi^2 = 6.2$) and $3\sigma$ (99.7\%, $\Delta\chi^2 = 18.4$) joint-confidence contours \citep{Press2002, BevingtonRobinson2003, HughesHase2010} for the $(P_0, \beta)$ parameter space. Similar results were obtained for the remaining two parameter combinations. Considering Dataset I we found the average period change to be $\dot{P} = 5.6 \times 10^{-11}~\textnormal{s}~\textnormal{s}^{-1}$. This value is about one order of magnitude smaller than the period decrease reported in \cite{Almeida2013}.

\subsection{Single companion model - Dataset I}

\begin{figure*}
\centering
\includegraphics[angle=0,scale=0.5]{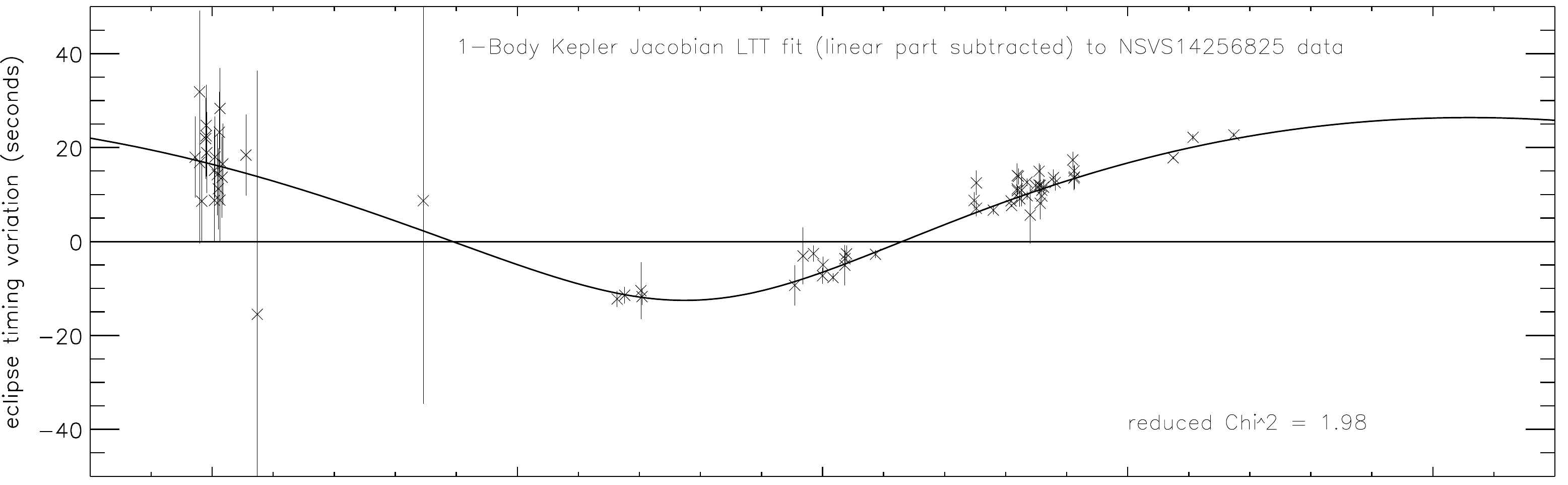}
\includegraphics[angle=0,scale=0.5]{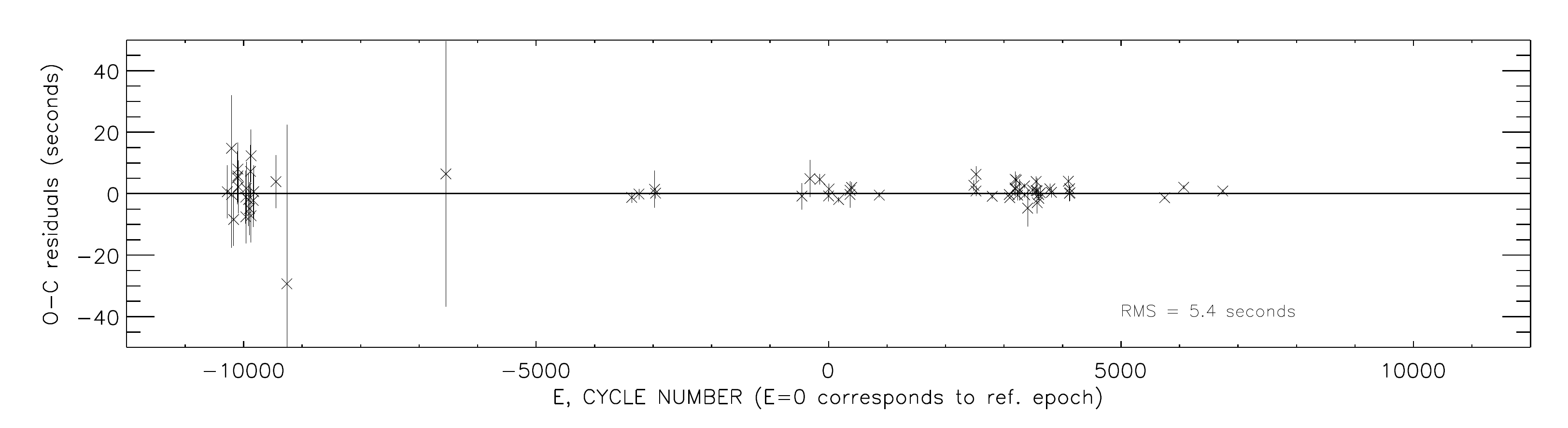}
\includegraphics[angle=0,scale=0.5]{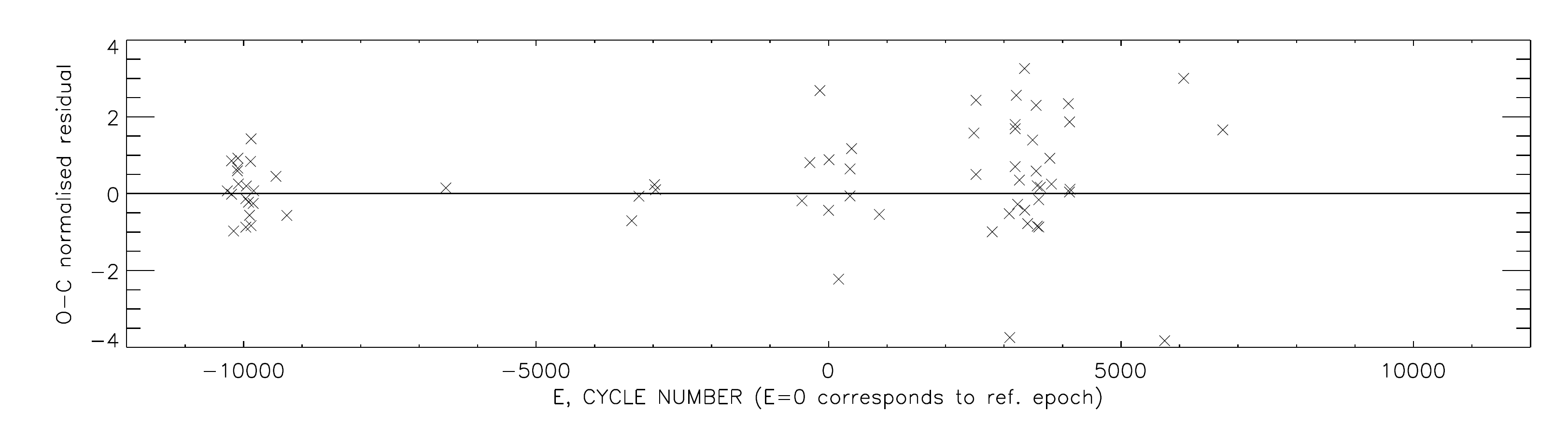}
\caption{Best-fit model (top panel) to the timing data as considered by \cite{Almeida2013}. We subtracted the linear part. \emph{Middle panel}: Residuals $O_{i}-C_{i}$ versus cycle number with a root-mean-square scatter (RMS) of about 5 seconds. \emph{Bottom panel}. Plot of normalised residuals $(O_{i}-C_{i})/\sigma_{i}$ (dimensionless, see text) between observed and 
computed times.} 
\label{1bodyLTTJacBestFit}
\end{figure*}

\begin{figure*}
\centering
\includegraphics[angle=0,scale=0.7]{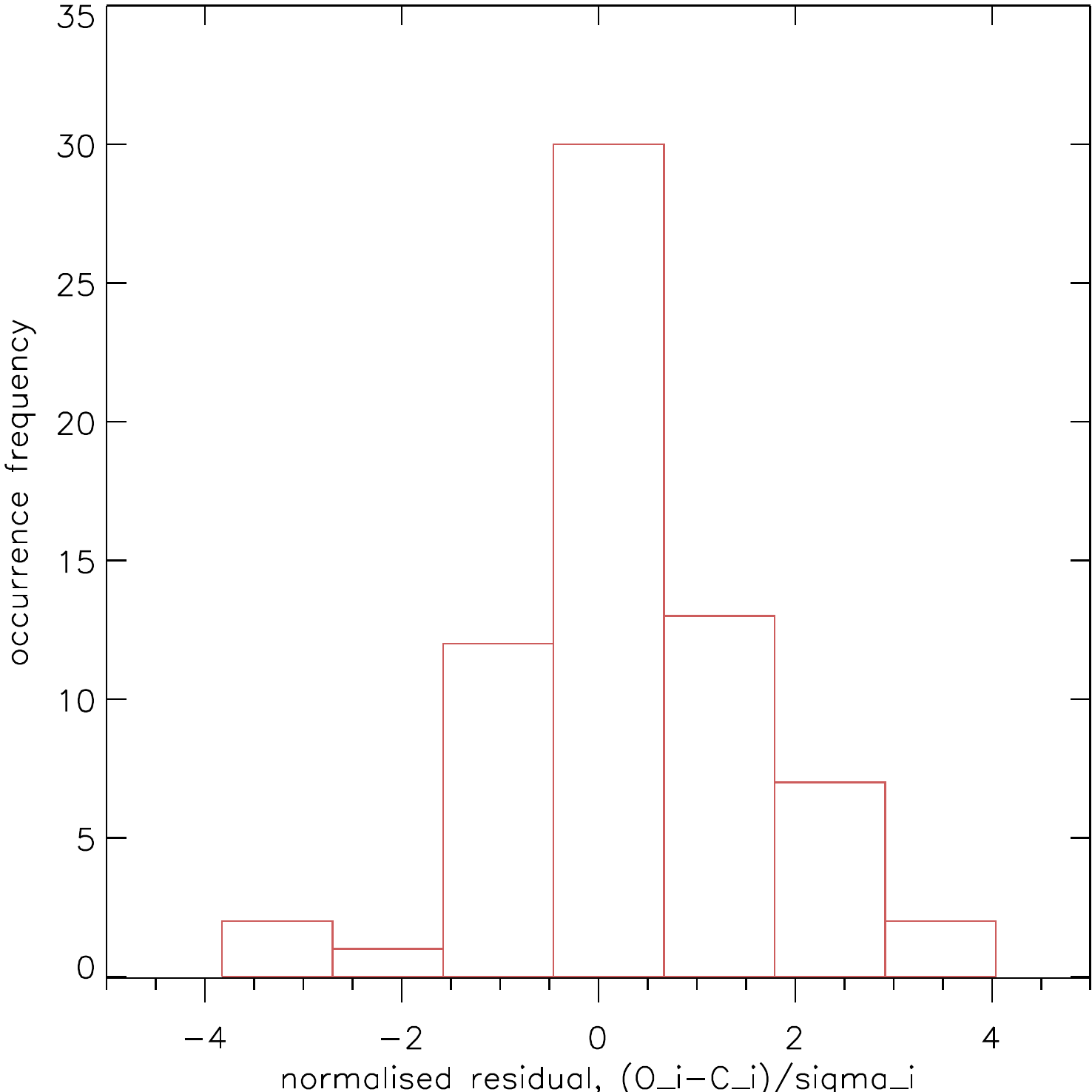}
\caption{Occurance frequency of normalised residuals (bottom panel in Fig.~\ref{1bodyLTTJacBestFit}) resembling a somewhat symmetric normal distribution. The units on the first axis are standard deviations with 1-$\sigma \simeq$ 5 seconds. The binsize was chosen to be 1.2$\sigma$.}
\label{histogram}
\end{figure*}

To reliably assess the validity of a two-companion model we first considered a one-companion model. Our period analysis yielded a shortest period of around $P_1 \simeq 7$ years (2557 days) with a semi-amplitude of $K_{1} \simeq 0.000231$ days (20 seconds). We therefore searched for a best-fit solution in a narrow interval around these values by seeding 523,110 initial guesses. The best-fit solution with $\chi^2_{60,0} = 1.98$ is shown in Fig.~\ref{1bodyLTTJacBestFit}. In Table \ref{bestfitparam} we show the corresponding best-fit parameters and derived quantities for the companion along with their formal (derived from the covariance matrix) $1\sigma$ uncertainties as obtained from \texttt{MPFIT}. Formal errors in the derived quantities were obtained from numerical error propagation, as described in \cite{BevingtonRobinson2003}. The residual plot in Fig.~\ref{1bodyLTTJacBestFit} (middle panel) shows no obvious trend above the 5 seconds level. The average timing uncertainty in the Almeida et al. (2013) data set is 5.5 seconds. An additional signal associated with a light-travel time effect should be detected on a $3\sigma$ level equivalent to a timing semi-amplitude of $\simeq 15$ seconds. Usually timing measurement are assumed to distribute normally around the expected model. We have therefore also plotted the normalised residuals $(O_{i}-C_{i})/\sigma_i$ \citep{HughesHase2010} as shown in the bottom panel of Fig.~\ref{1bodyLTTJacBestFit}. The corresponding histogram is shown in Fig.~\ref{histogram}. Whether the timing residuals follow a Gaussian distribution is unclear at the moment.

\begin{figure*}
\centering
\includegraphics[angle=0,scale=0.45]{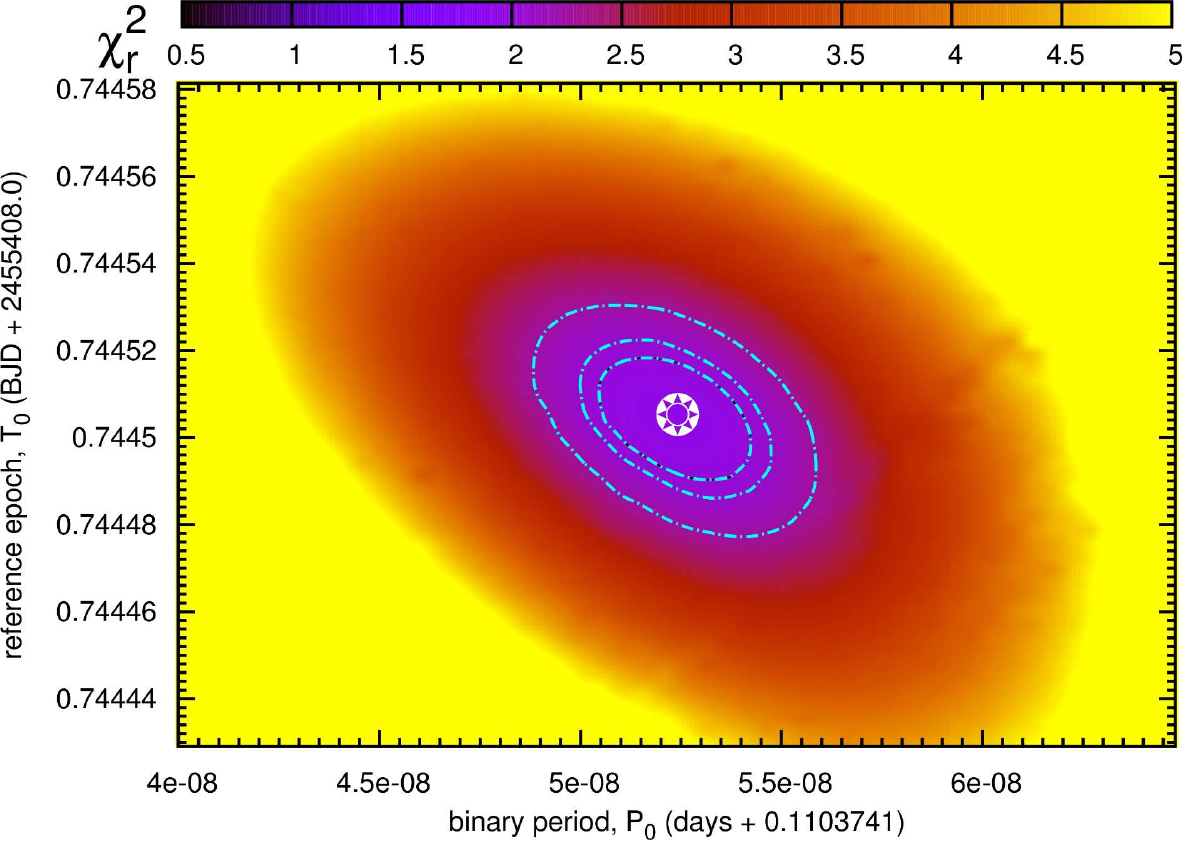}
\includegraphics[angle=0,scale=0.45]{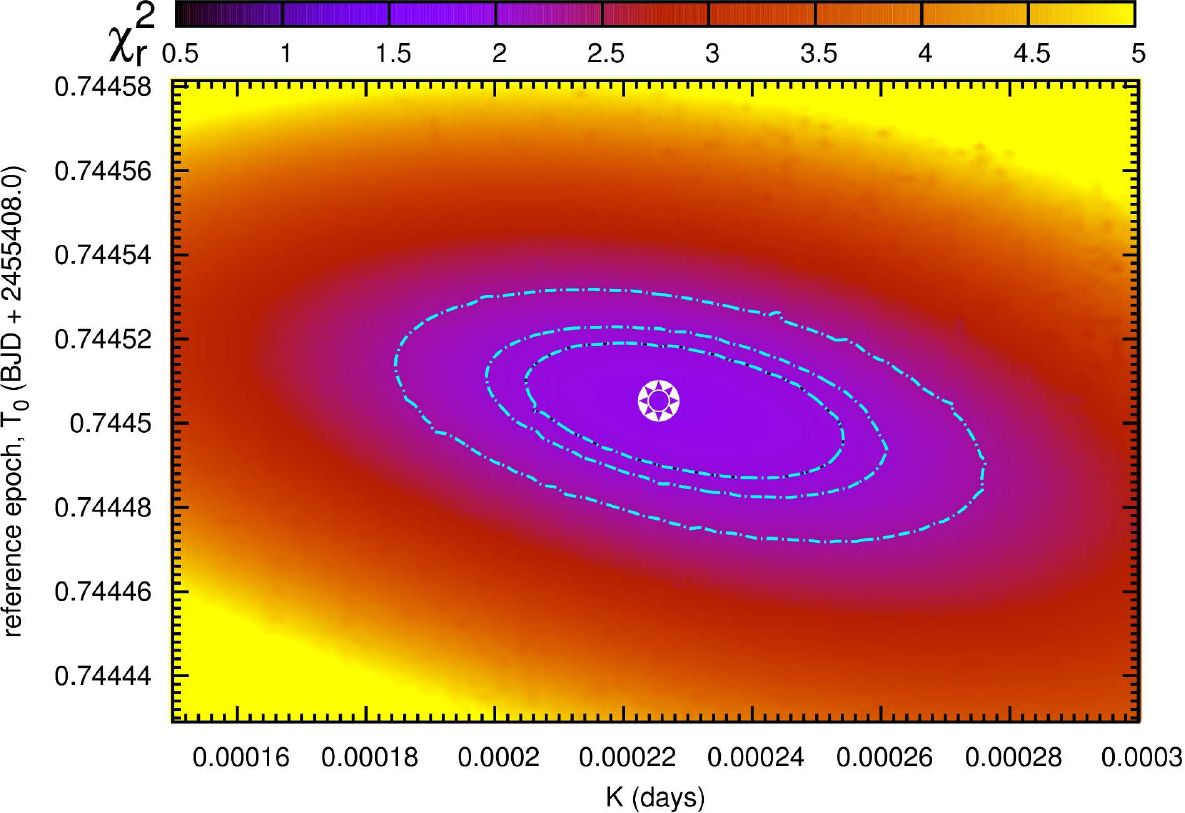}
\includegraphics[angle=0,scale=0.45]{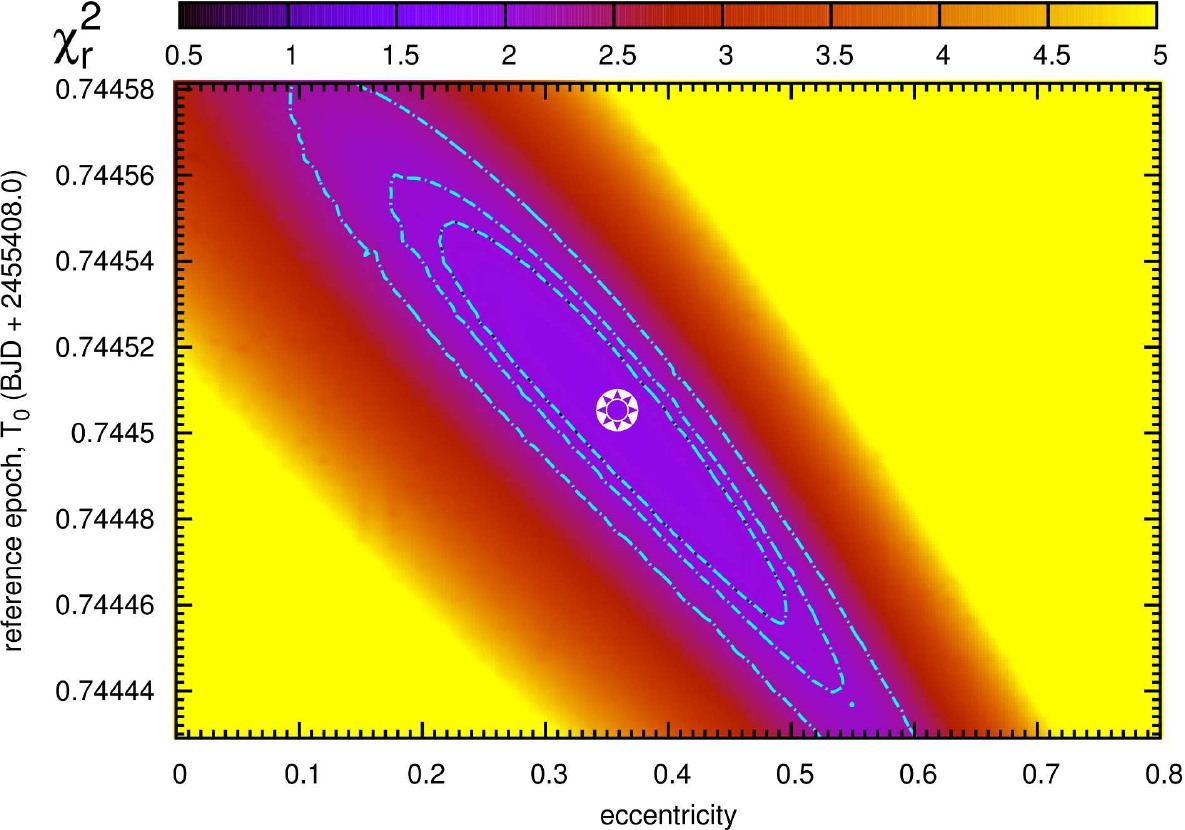}

\includegraphics[angle=0,scale=0.45]{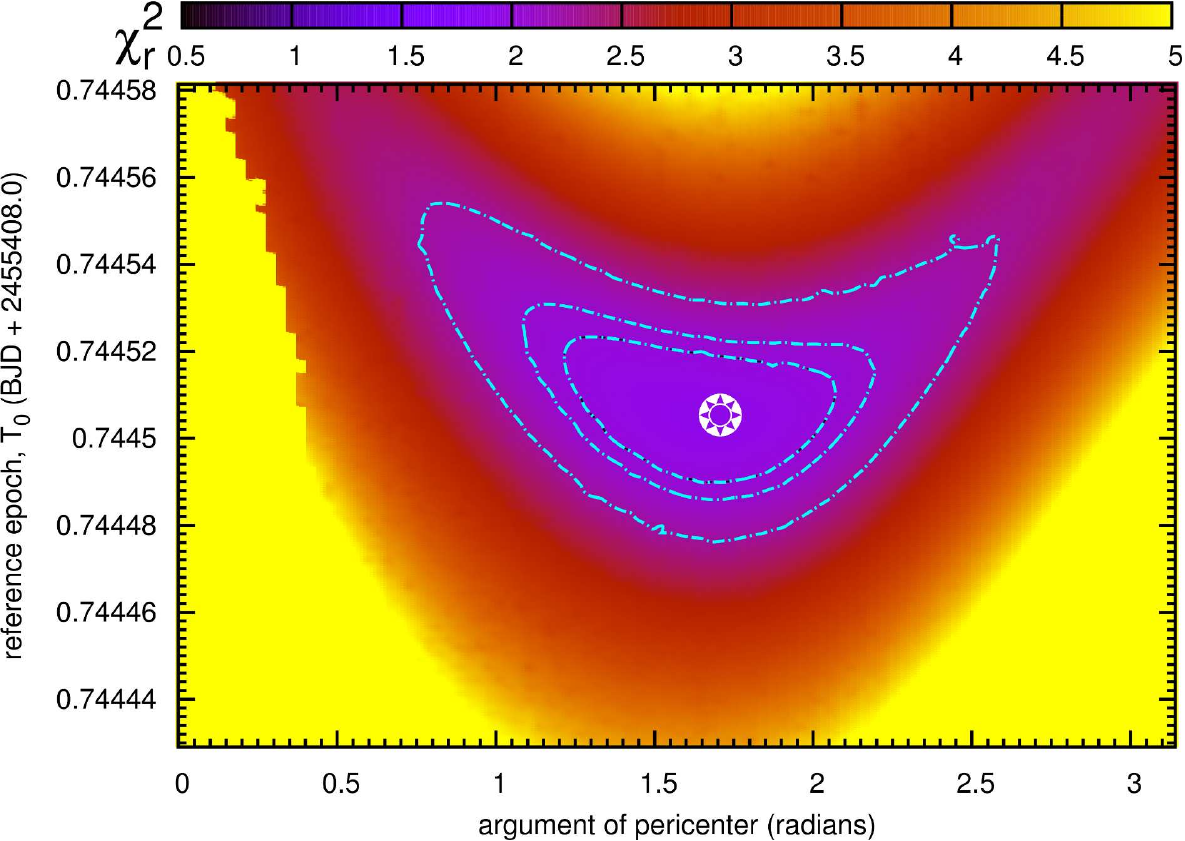}
\includegraphics[angle=0,scale=0.45]{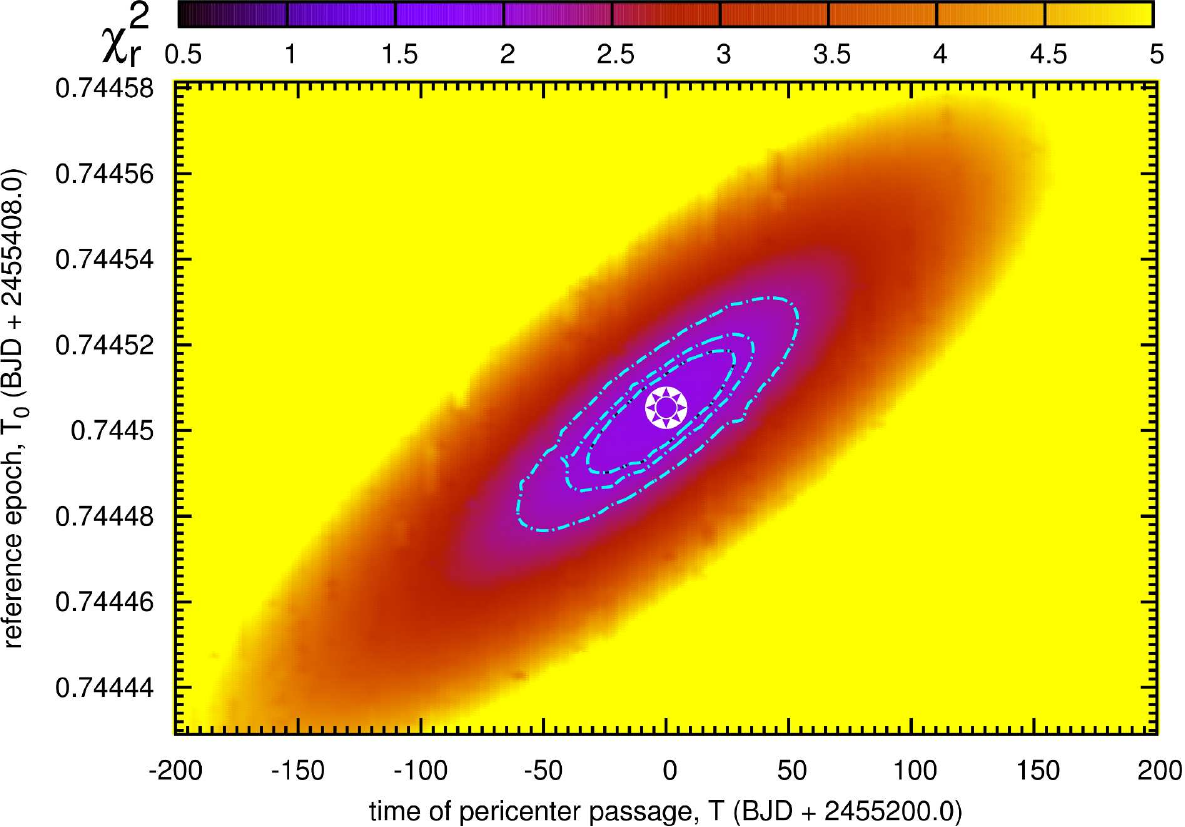}
\includegraphics[angle=0,scale=0.45]{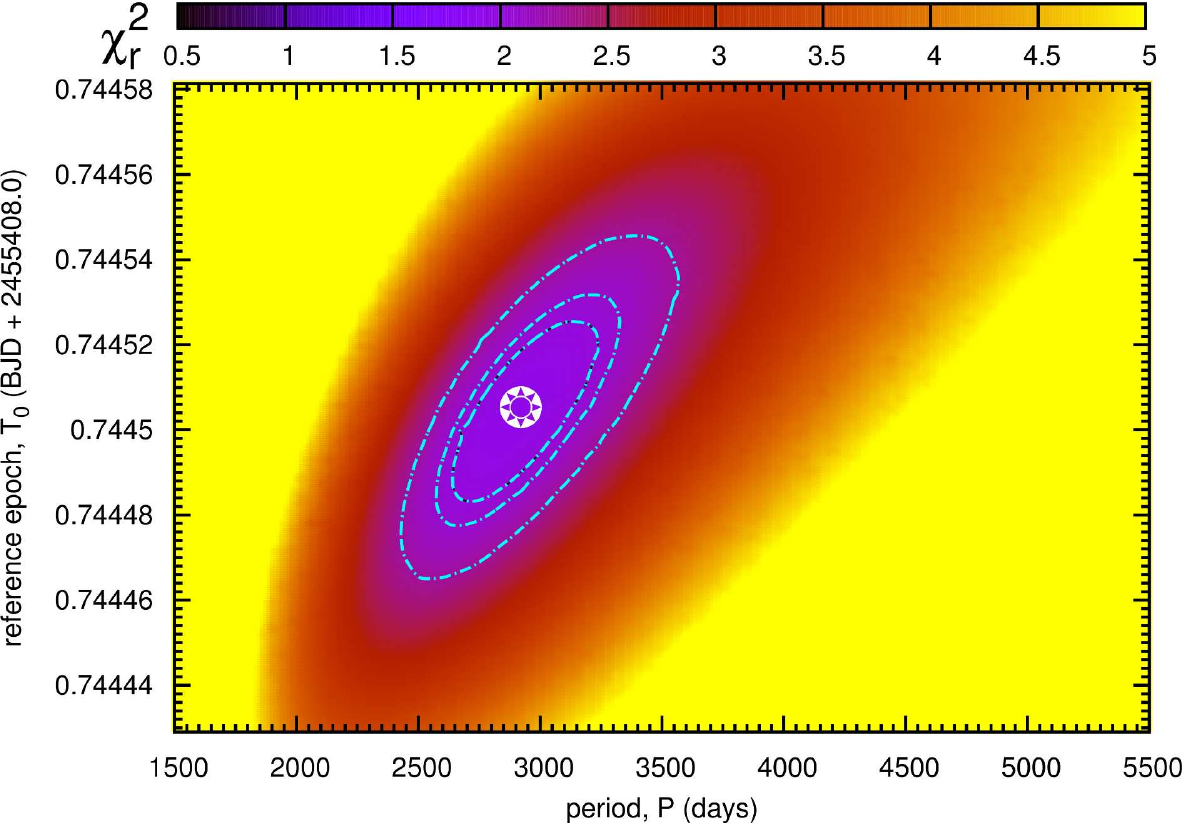}

\includegraphics[angle=0,scale=0.45]{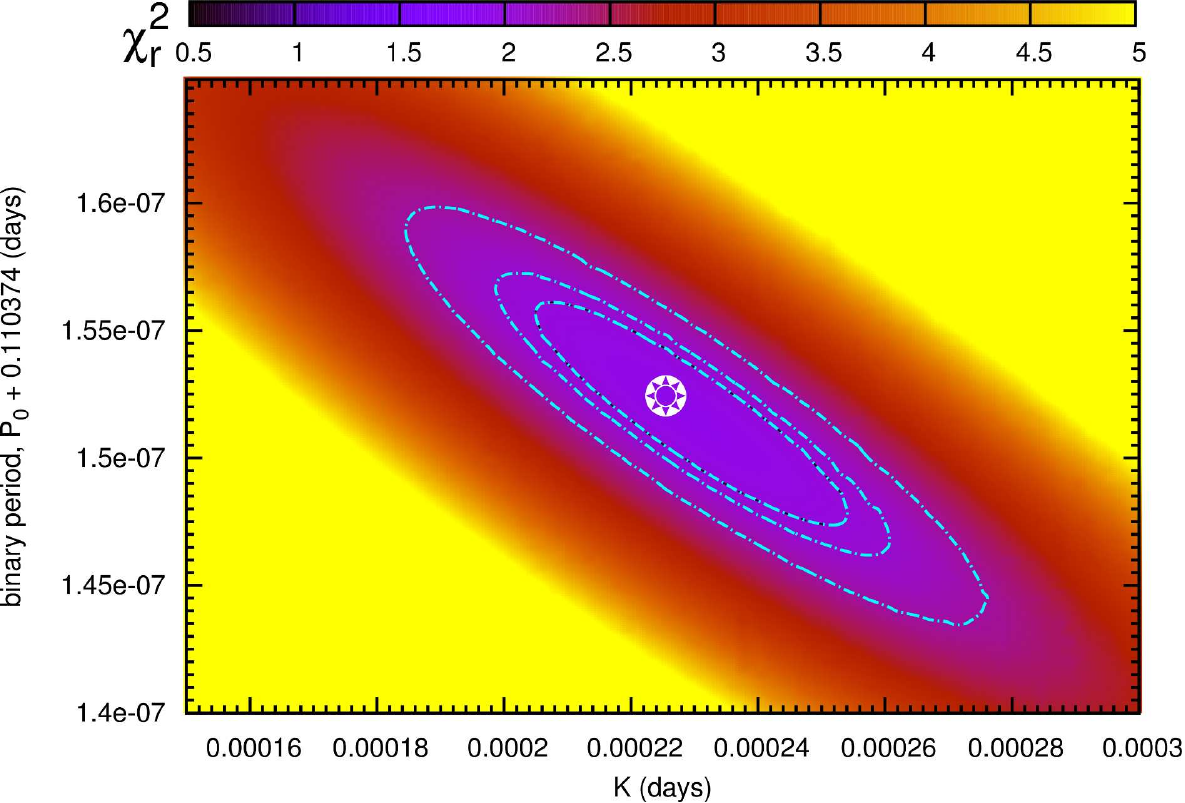}
\includegraphics[angle=0,scale=0.45]{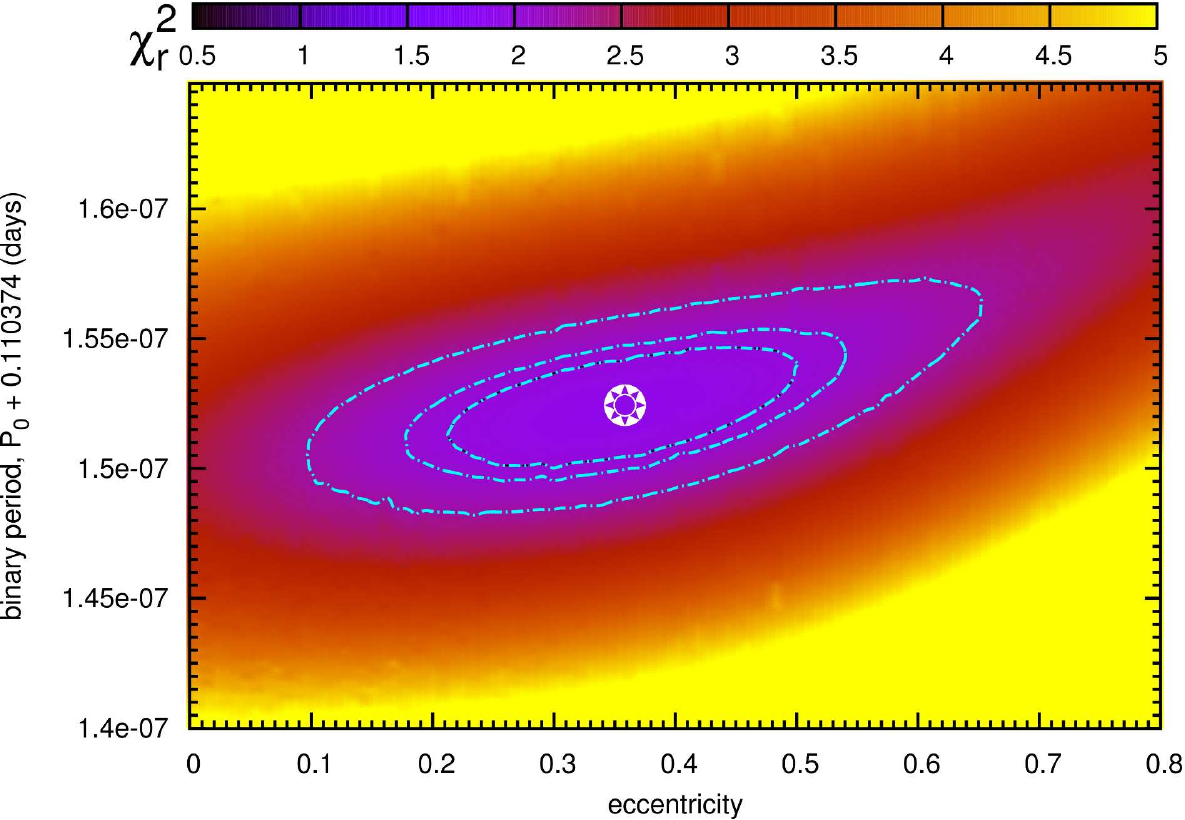}
\includegraphics[angle=0,scale=0.45]{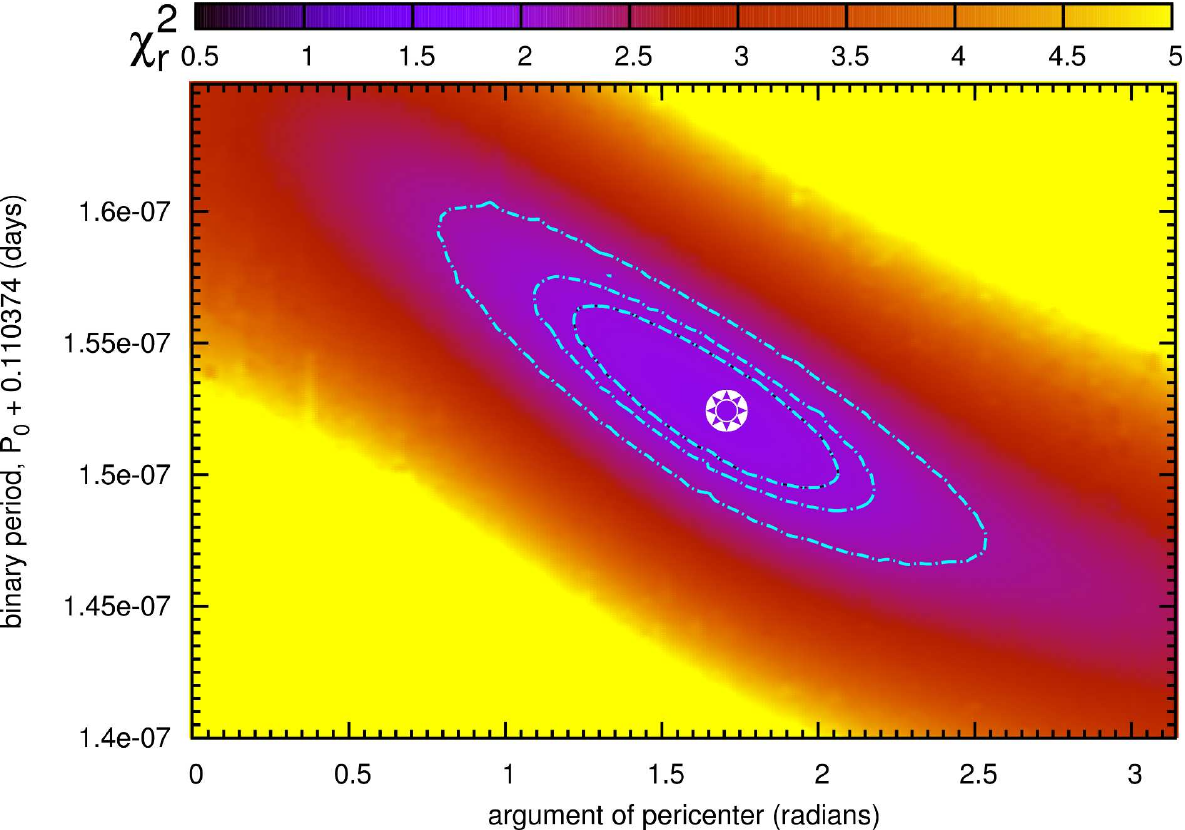}

\includegraphics[angle=0,scale=0.45]{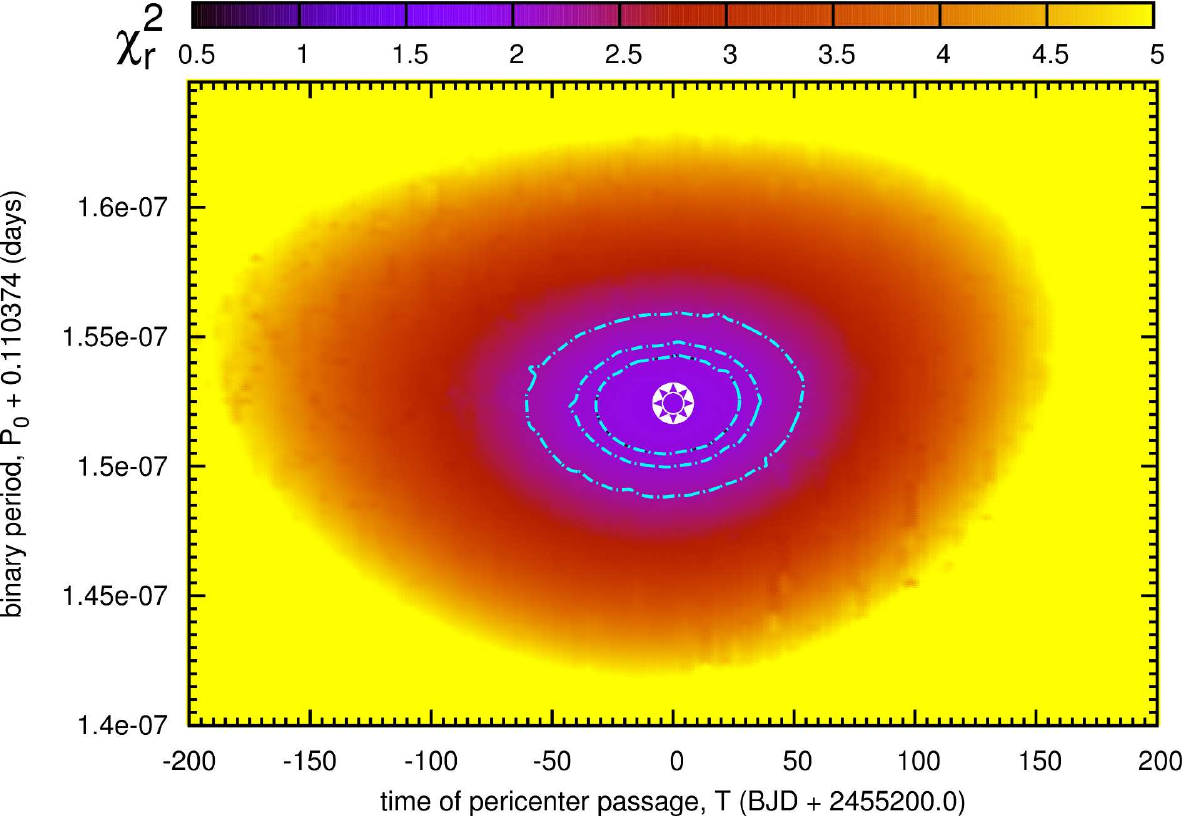}
\includegraphics[angle=0,scale=0.45]{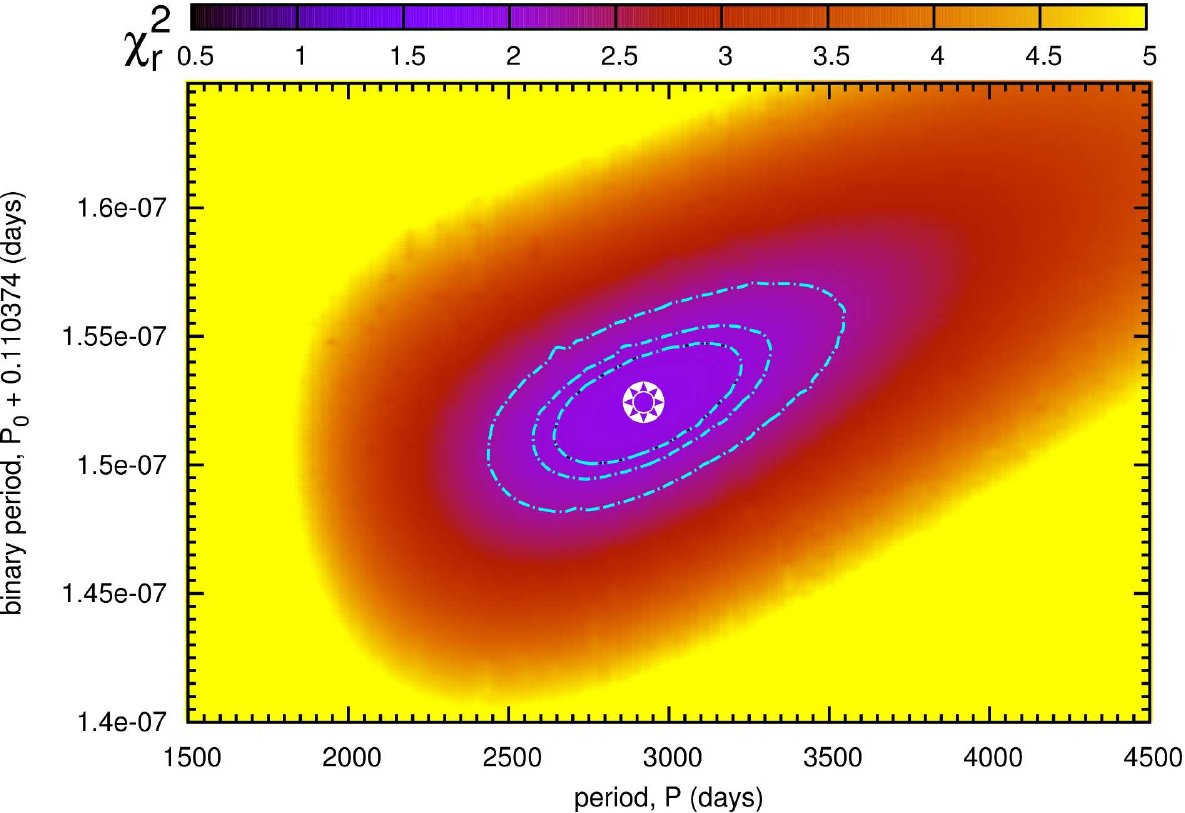}
\includegraphics[angle=0,scale=0.45]{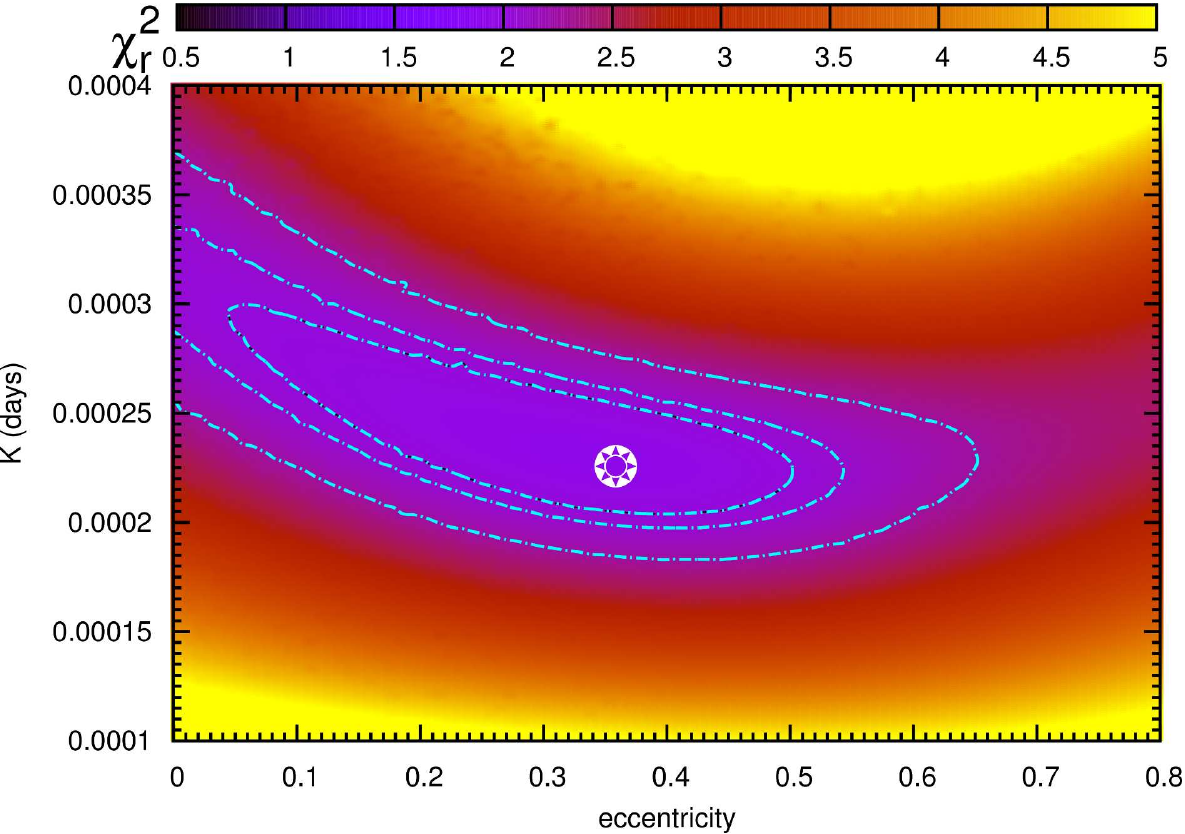}

\includegraphics[angle=0,scale=0.45]{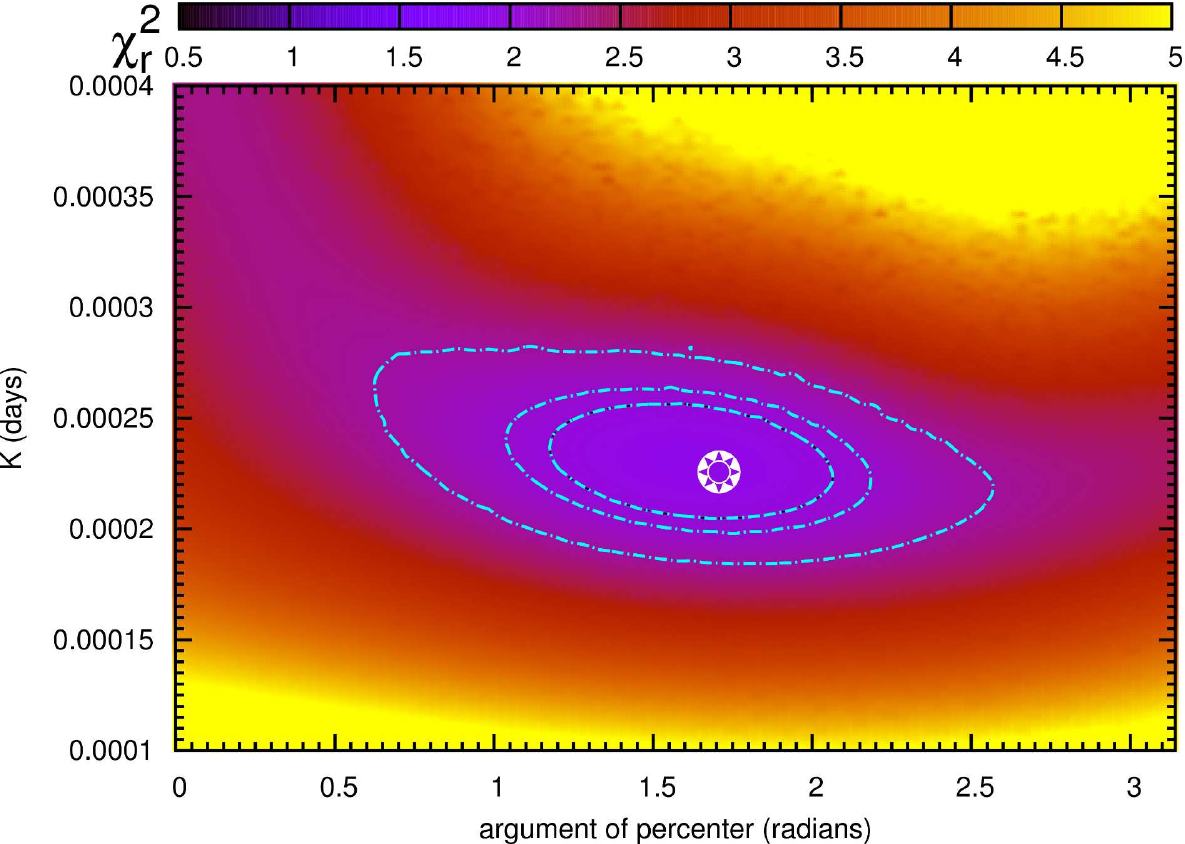}
\includegraphics[angle=0,scale=0.45]{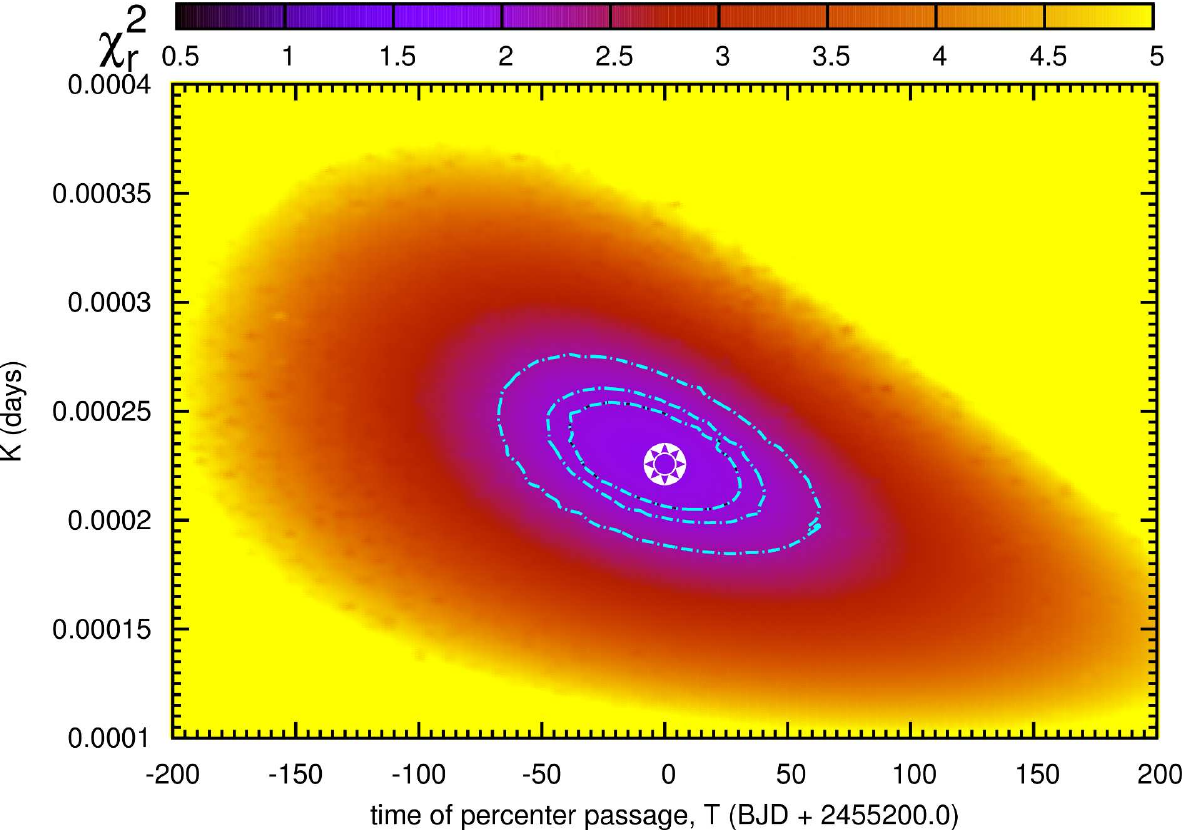}
\includegraphics[angle=0,scale=0.45]{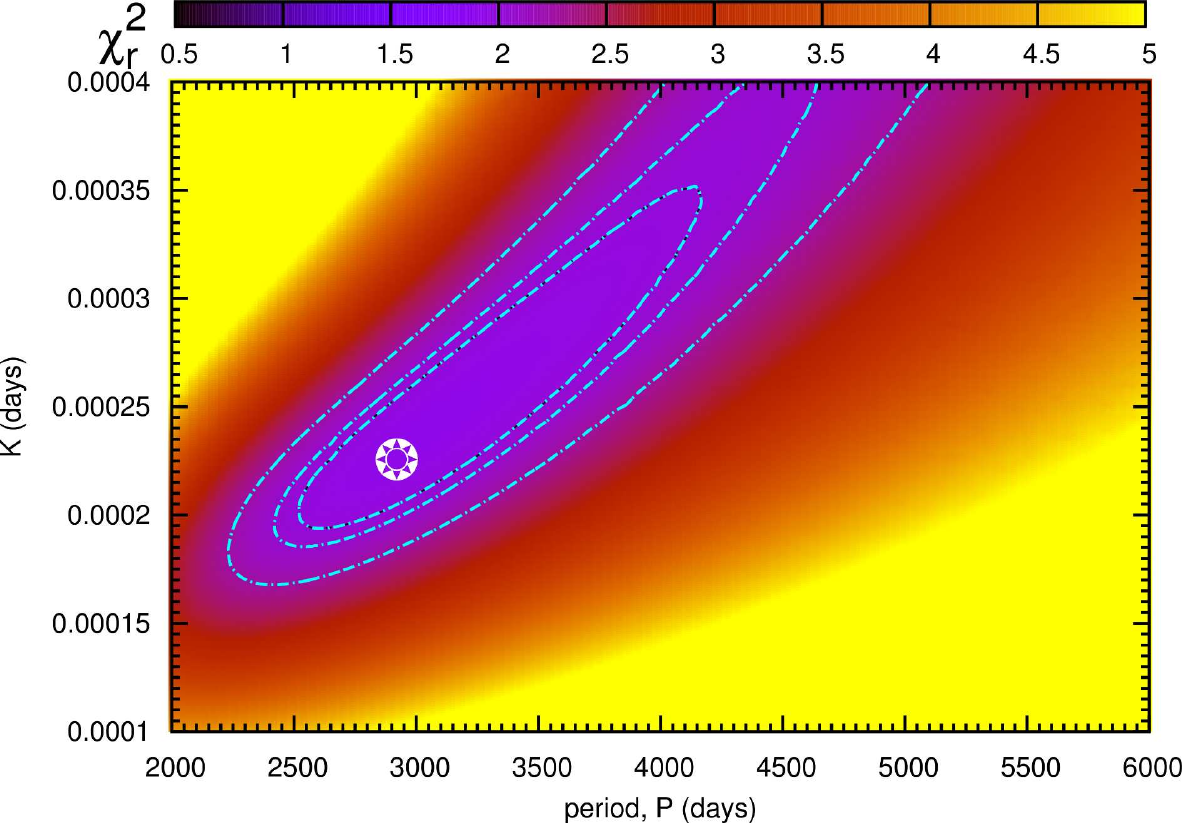}
\caption{Similar to Fig.~\ref{2DMapSecularBestFitModel} but showing scans of orbital parameters for the linear + one-LTT model. {\it See electronic version for color figures}.} 
\label{2DScansFig1}
\end{figure*}

\begin{figure*}
\centering
\includegraphics[angle=0,scale=0.45]{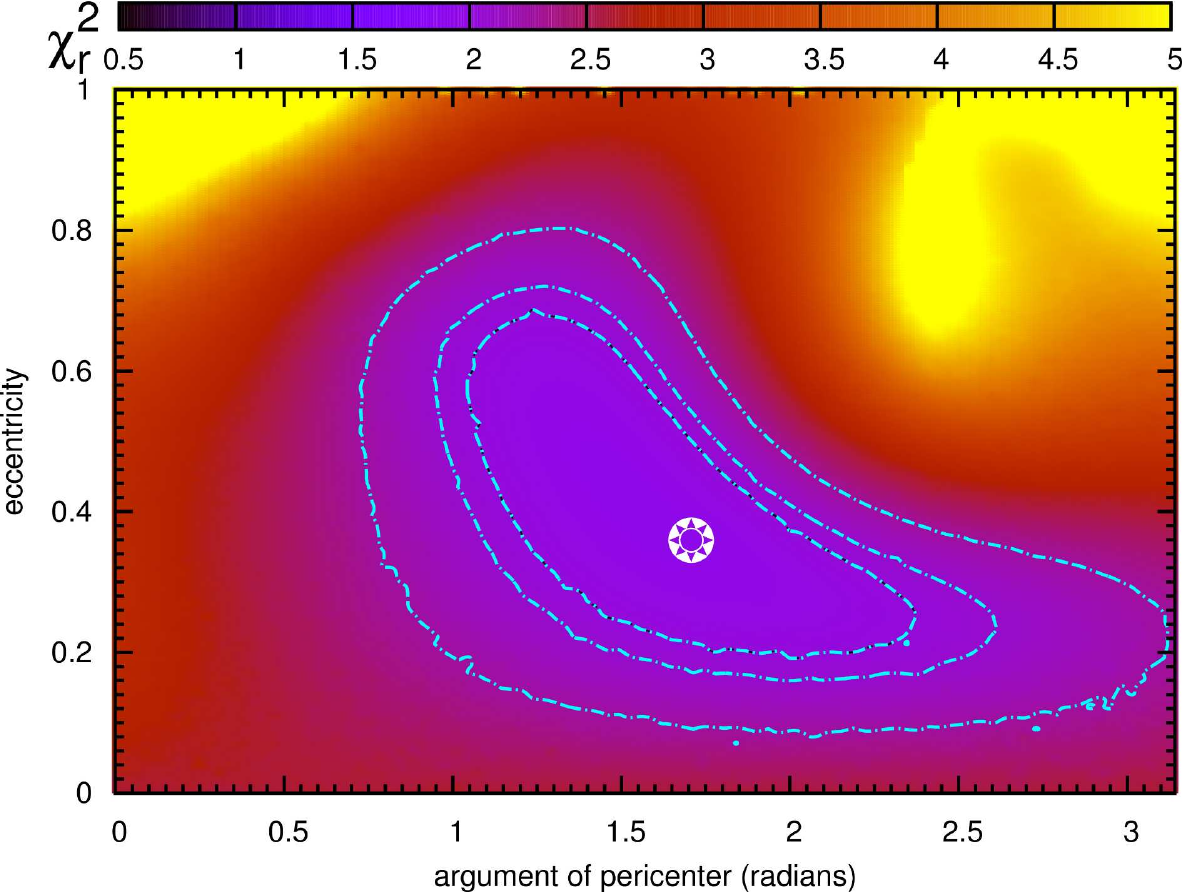}
\includegraphics[angle=0,scale=0.45]{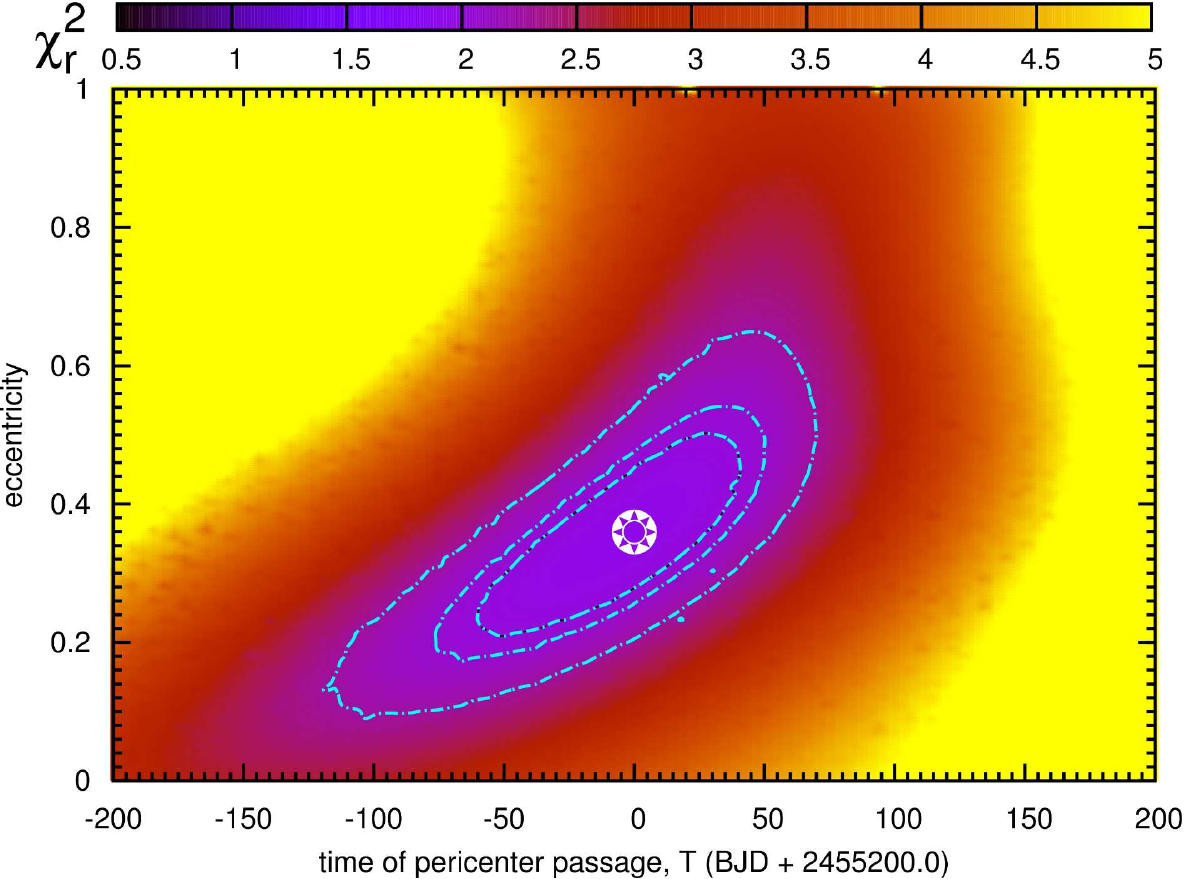}
\includegraphics[angle=0,scale=0.45]{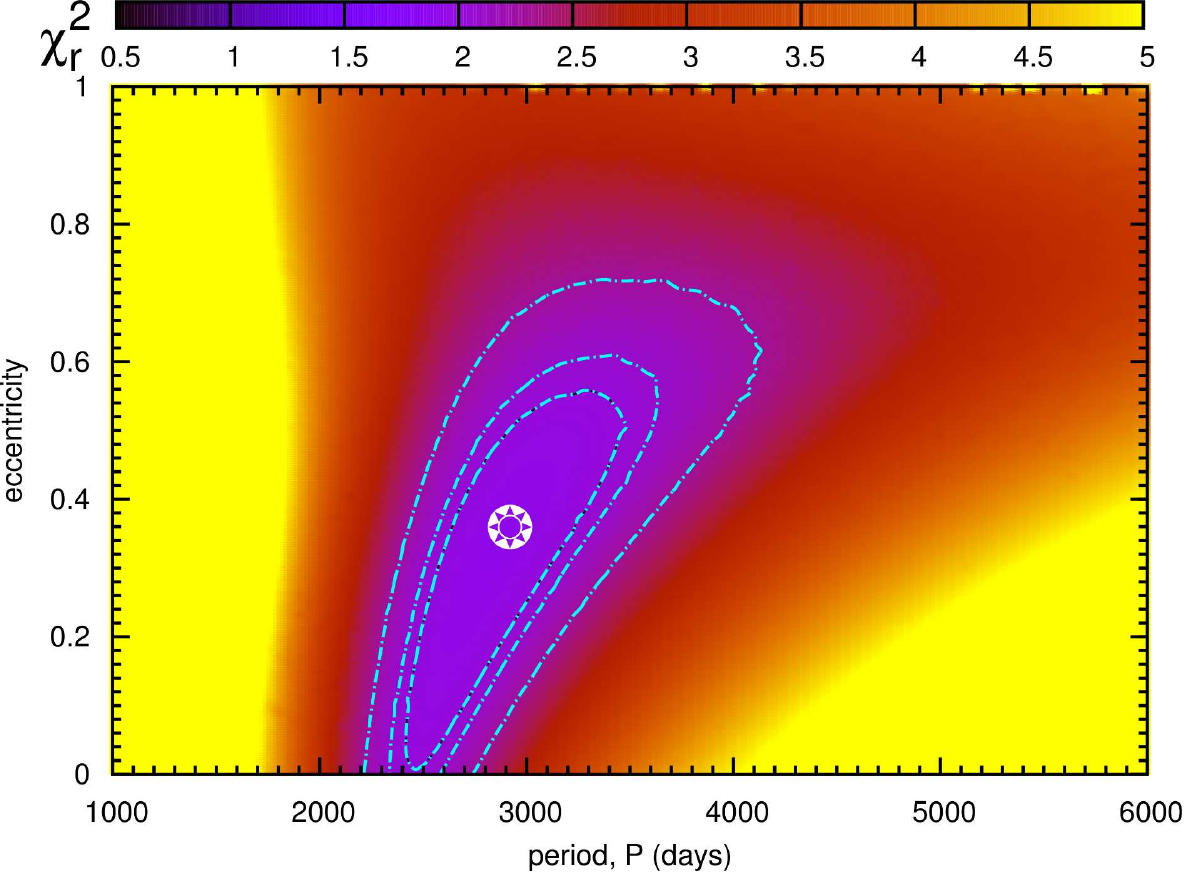}

\includegraphics[angle=0,scale=0.45]{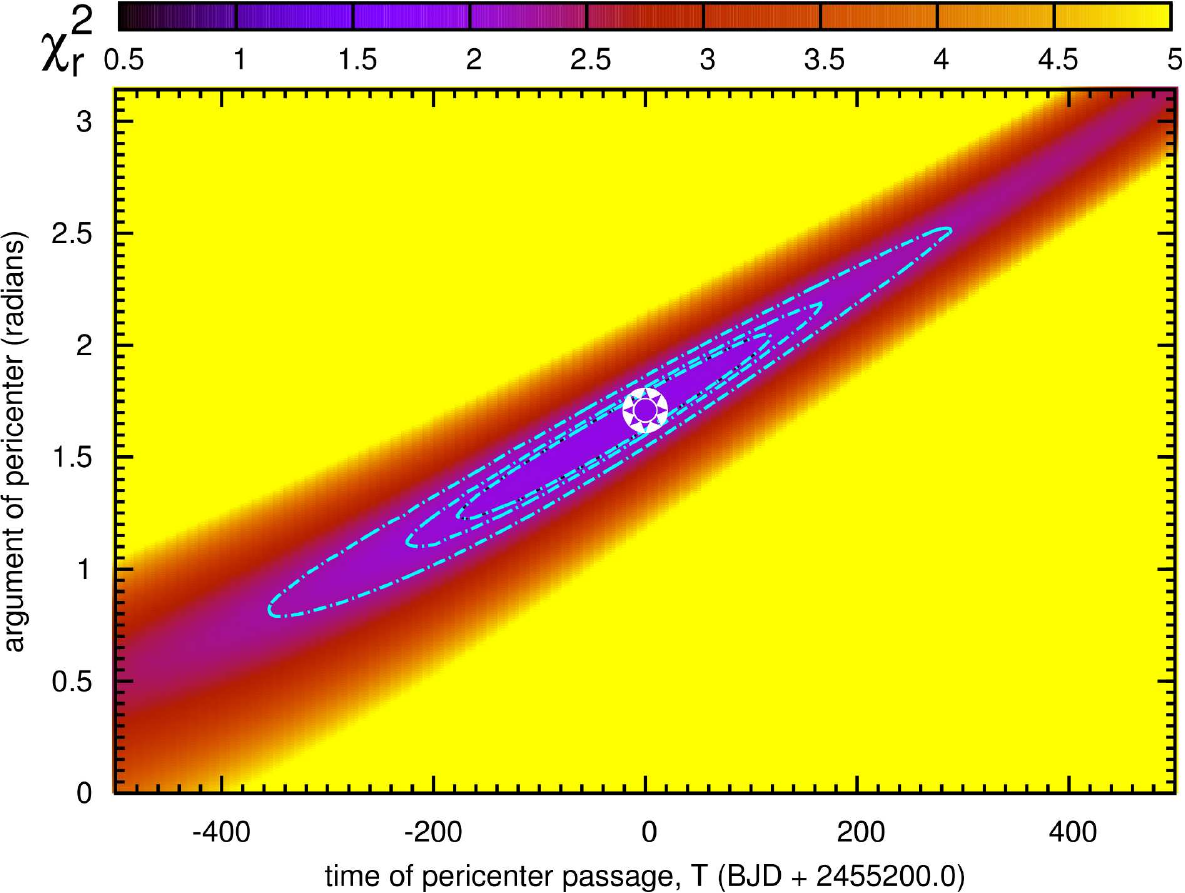}
\includegraphics[angle=0,scale=0.45]{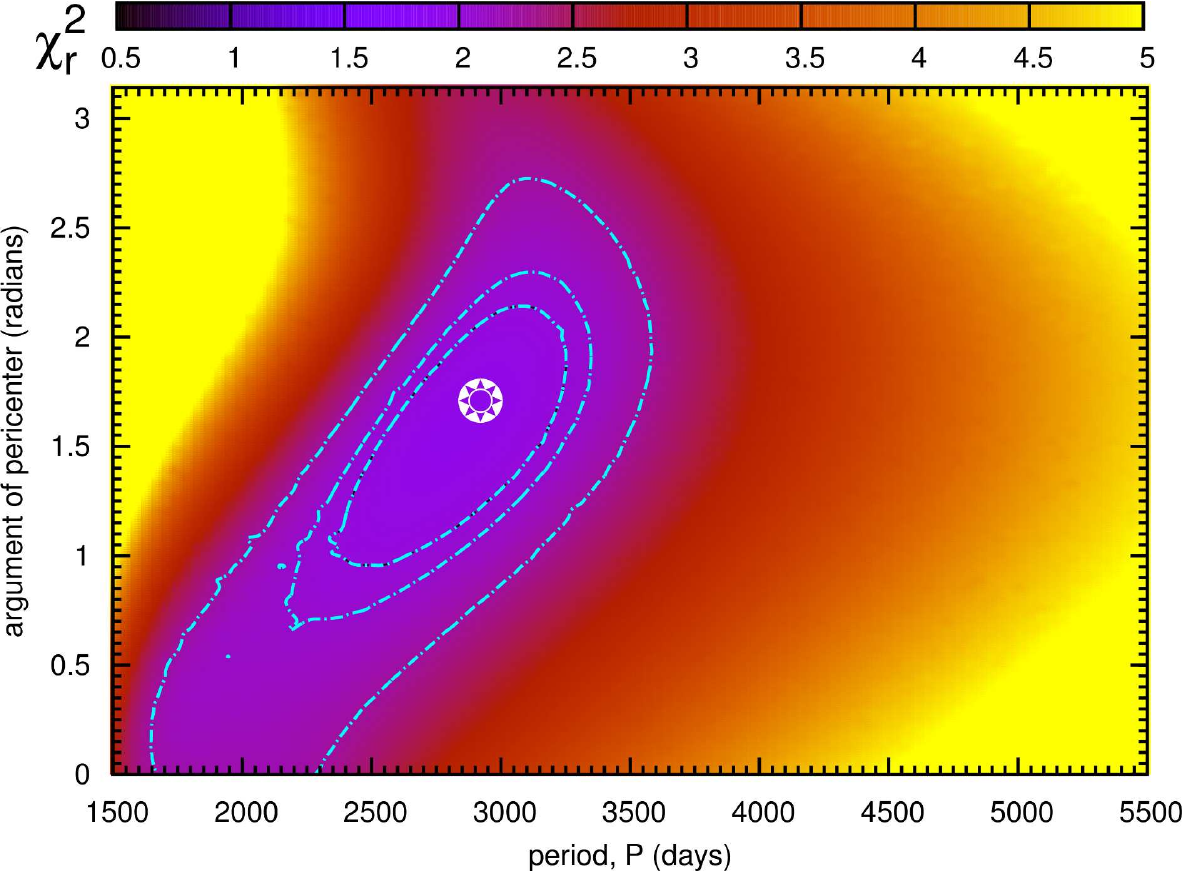}
\includegraphics[angle=0,scale=0.45]{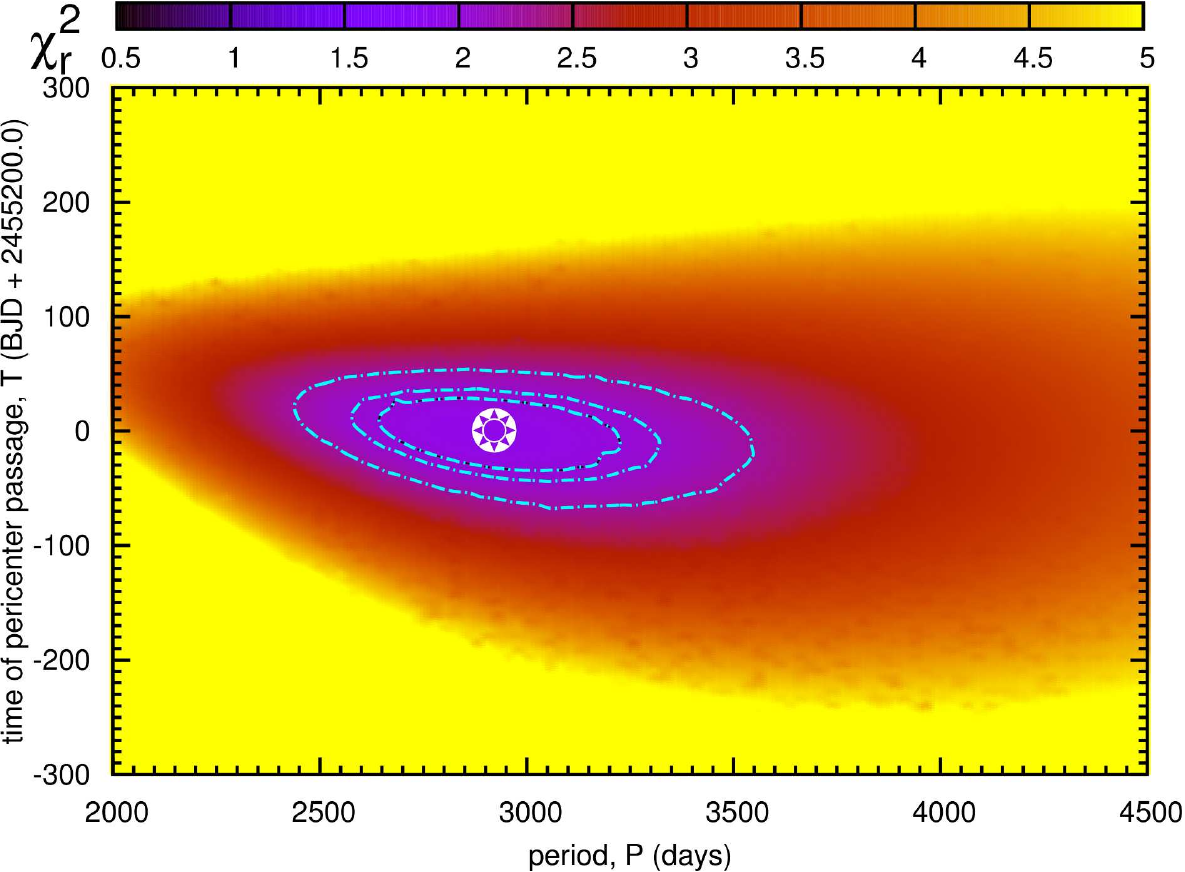}
\caption{Similar to Fig.~\ref{2DScansFig1} but for the remaining six parameter combinations. {\it See electronic version for colors}.} 
\label{2DScansFig2}
\end{figure*}

Again, we have explored the $\chi^2_{60}$ function in the vicinity of the best-fit parameters and determined two-dimensional joint-confidence intervals. We show all 21 two-parameter combinations in Fig.~\ref{2DScansFig1} and Fig.~\ref{2DScansFig2}. While the two considered parameters in a given panel were kept fixed, we allowed all the remaining parameters to re-optimise (with an initial guess given by the best-fit values listed in Table \ref{bestfitparam}) during a LM iteration \citep{BevingtonRobinson2003}.

We note that several of the parameters correlate with each other. This is especially true for the $(T_1,\omega_1)$ pair shown in Fig.~\ref{2DScansFig2}. Choosing our reference epoch $T_0$ to be close to the middle of the data set results in almost no correlation between $T_0$ and $P_0$ (see top left panel in Fig.~\ref{2DScansFig1}). In addition, we note that the $\chi^2$ topology around the best-fit parameters deviates from its expected parabolic form. This is most readily apparent in the $(\omega_1,e_1)$ panel in Fig.~\ref{2DScansFig2}.

\begin{figure*}
\centering
\includegraphics[angle=0,scale=1.0]{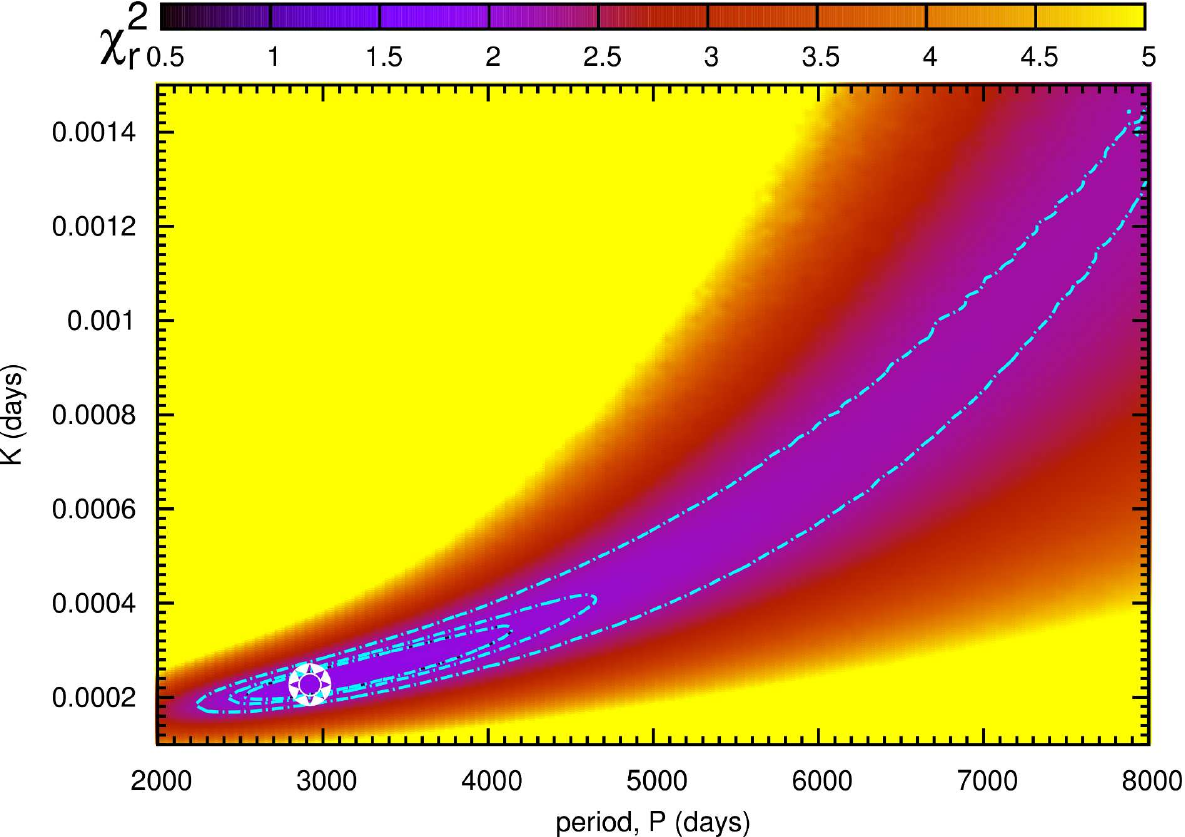}
\caption{Best-fit solution and the result of calculating the two-dimensional joint-confidence contours of a single-companion model over a larger parameter interval. {\it See electronic version for colors}.} 
\label{2DScansFig16Zoom}
\end{figure*}

Finally, we note that the $3\sigma$ confidence level in the $(P_{1},K_{1})$ (bottom-right) panel of Fig.~\ref{2DScansFig1} appears open, and stretches toward longer periods $(P_1)$ and larger semi-amplitudes ($K_1$). With this in mind, we then recalculated the $\chi^2$ space of $(P_{1},K_{1})$ considering a larger interval in the two parameters. The result is shown in Fig.~\ref{2DScansFig16Zoom}, demonstrating that the 3-sigma joint-confidence contour remains open for orbital periods larger than around 22 years. We therefore suspect that our best-fit model resides within a local minimum.

\begin{figure*}
\centering
\includegraphics[angle=0,scale=0.7]{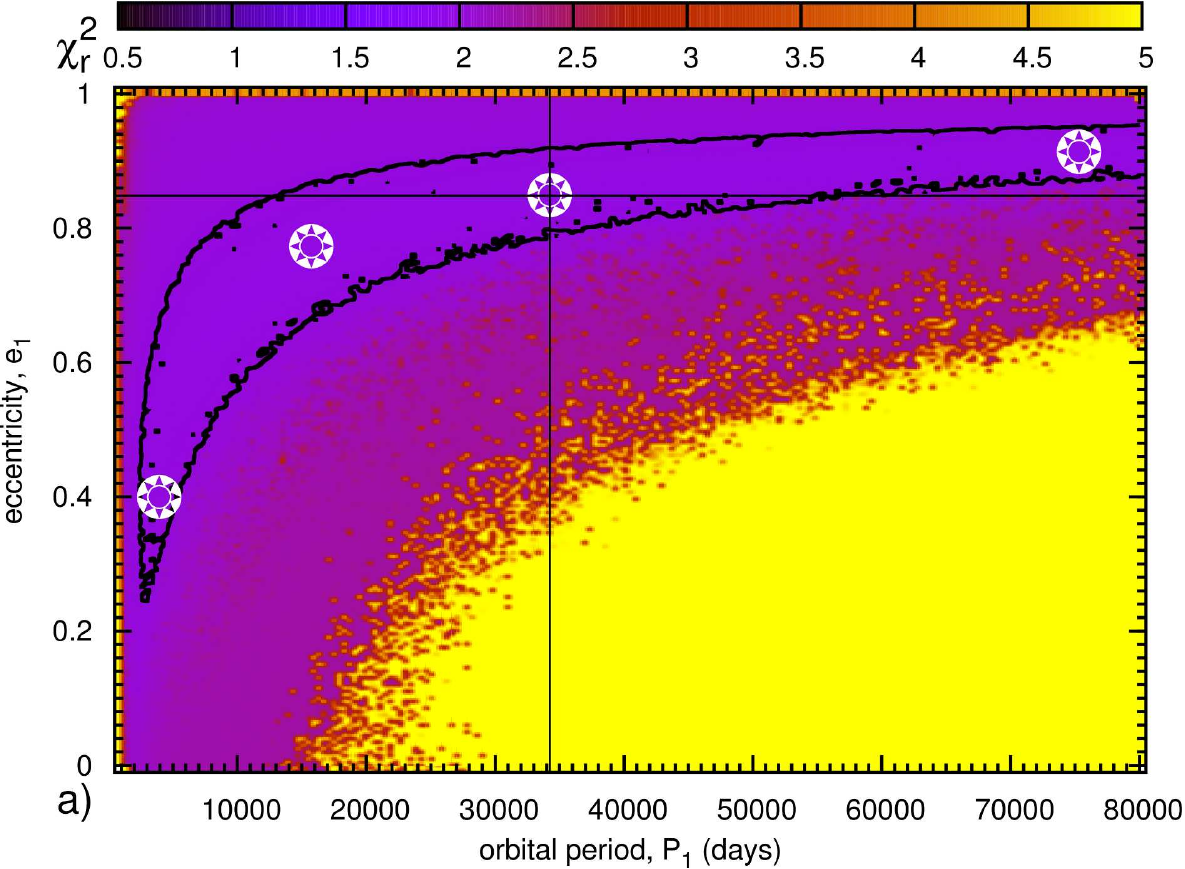}
\includegraphics[angle=0,scale=0.7]{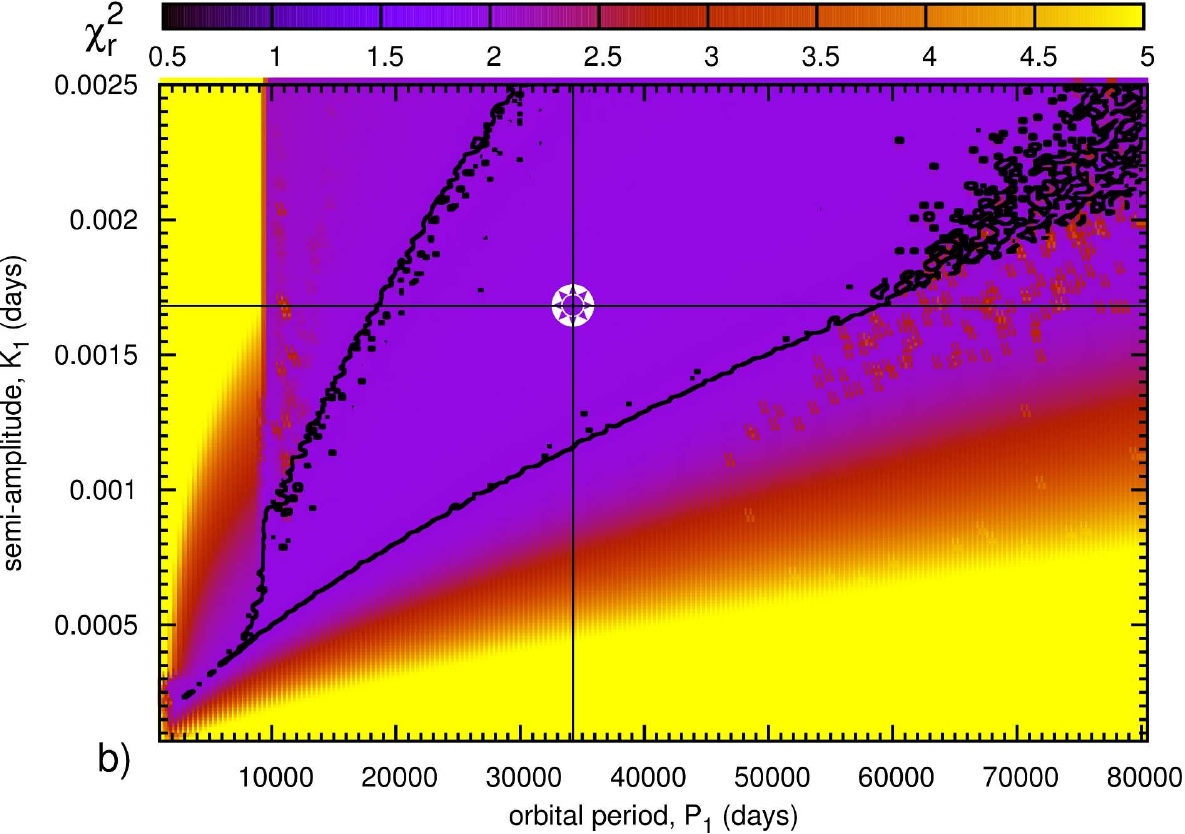}
\includegraphics[angle=0,scale=0.7]{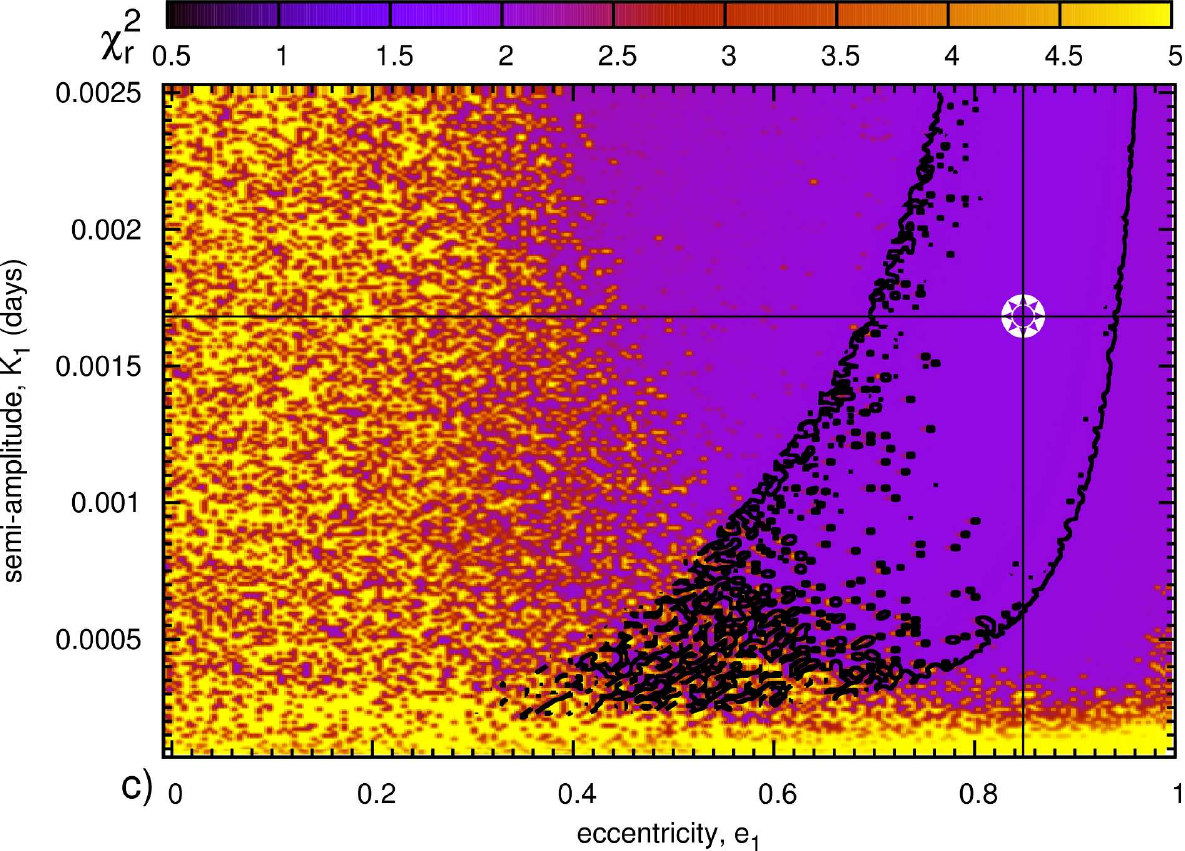}
\caption{Best-fit solution (cross-hair) when searching over a large search grid in the period and semi-amplitude. We show the result of calculating the two-dimensional joint-confidence contour with $\Delta\chi^2 = 2.3$ (68.3\%). The black contour line is the 1-sigma confidence level with $\chi^2_{60,0}=1.993$. In panel a) we show also three randomly chosen pairs of $(P_1,e_1)$ all within the 1-sigma level. {\it See electronic version for colors}.} 
\label{2DScansThreeFigures}
\end{figure*}

To test whether we are dealing with a local minimum we explore the $\chi^2$ parameter space on a wider search grid by following the approach as outlined previously. Surprisingly, we found a marginally improved solution with a 
smaller best-fit $\chi^2_{60,0*}$ value of 1.96, a reduction by $2\%$ compared to the first best-fit solution of 1.98. Computing the $\chi^2_{60}$ space 
around the new best-fit solution over a large interval in the parameters $K_1, e_1$ and $P_1$ resulted in Fig.~\ref{2DScansThreeFigures}. 

In each panel, our (new) improved best-fit solution is marked by a cross-hair. The corresponding model parameters are shown in Table \ref{bestfitparam_extended}. We have omitted quoting the formal uncertainties for reasons that will become apparent shortly. In Fig.~\ref{2DScansThreeFigures} we also show the 1-sigma (68.3\%) joint-confidence contour of $\Delta\chi^2_{60} = 1.993$ (black line) encompassing our best-fit model. Our results suggest that a plethora of models, with $\chi^2_{60}$ within the 1-sigma confidence level, are equally capable of explaining the timing data. Statistically, within the 1-sigma uncertainty region, essentially no differences in the $\chi^2$ exist between the various solutions.

\begin{figure*}
\centering
\includegraphics[angle=0,scale=0.25]{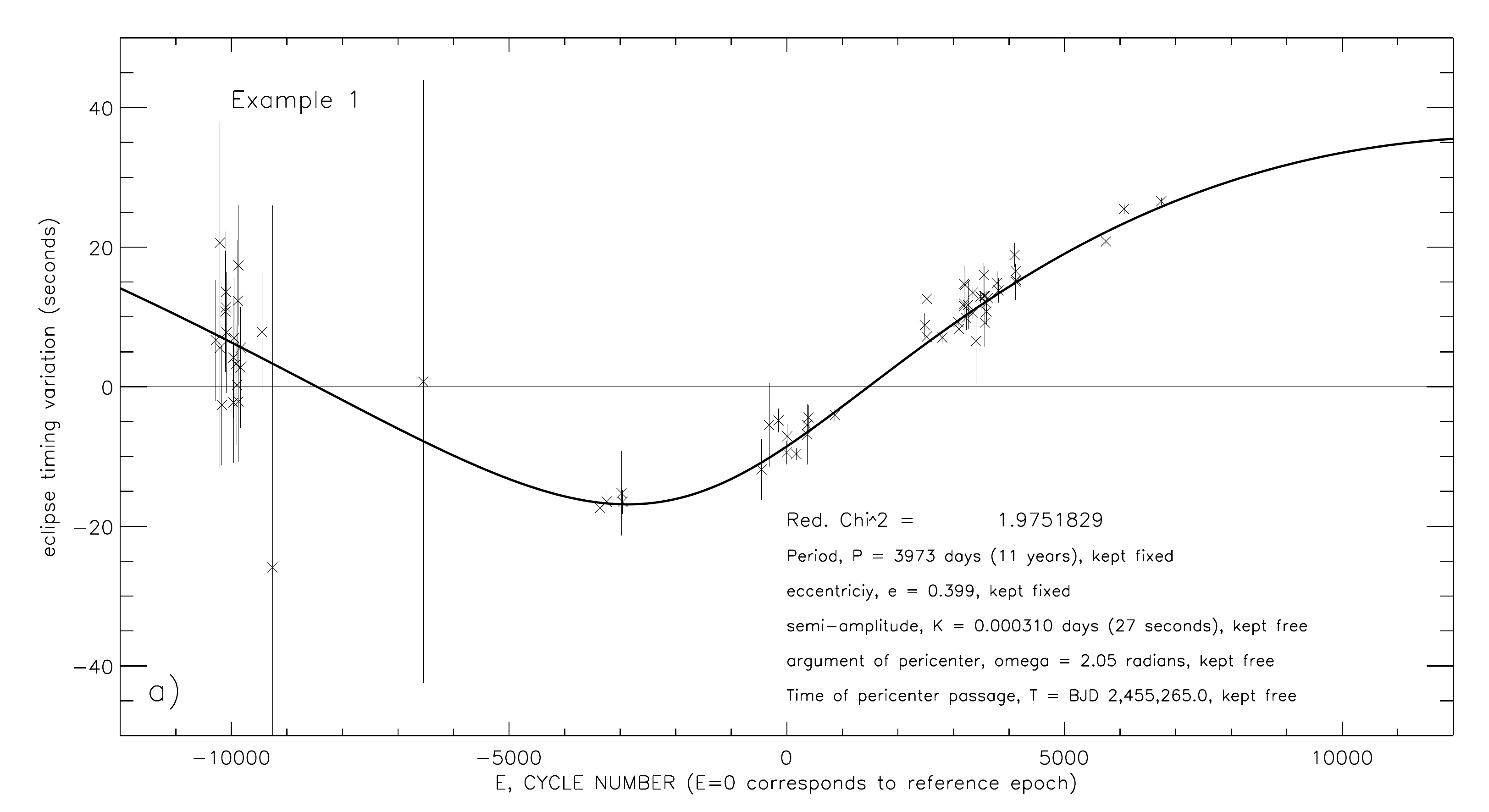}
\includegraphics[angle=0,scale=0.25]{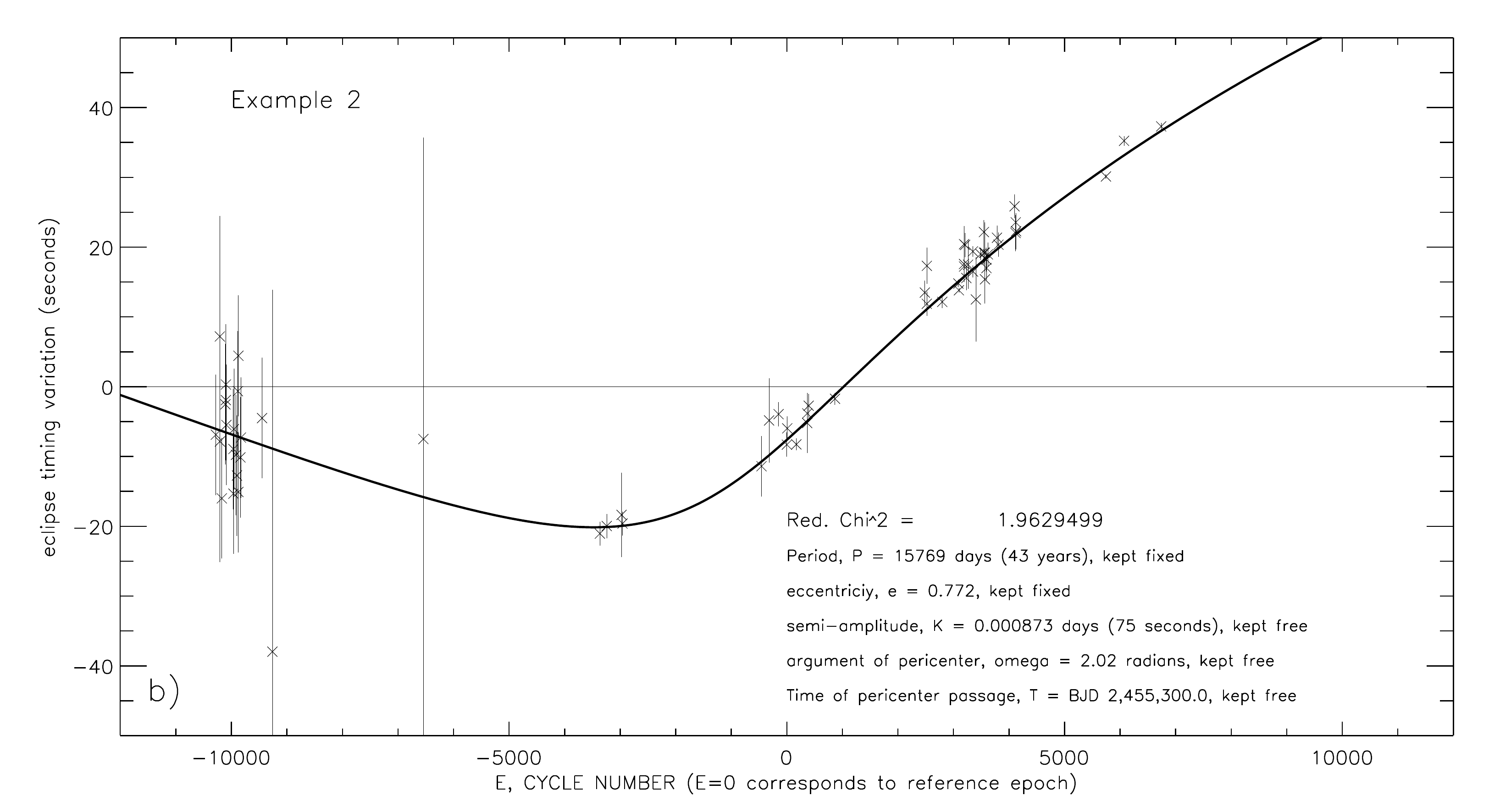}
\includegraphics[angle=0,scale=0.25]{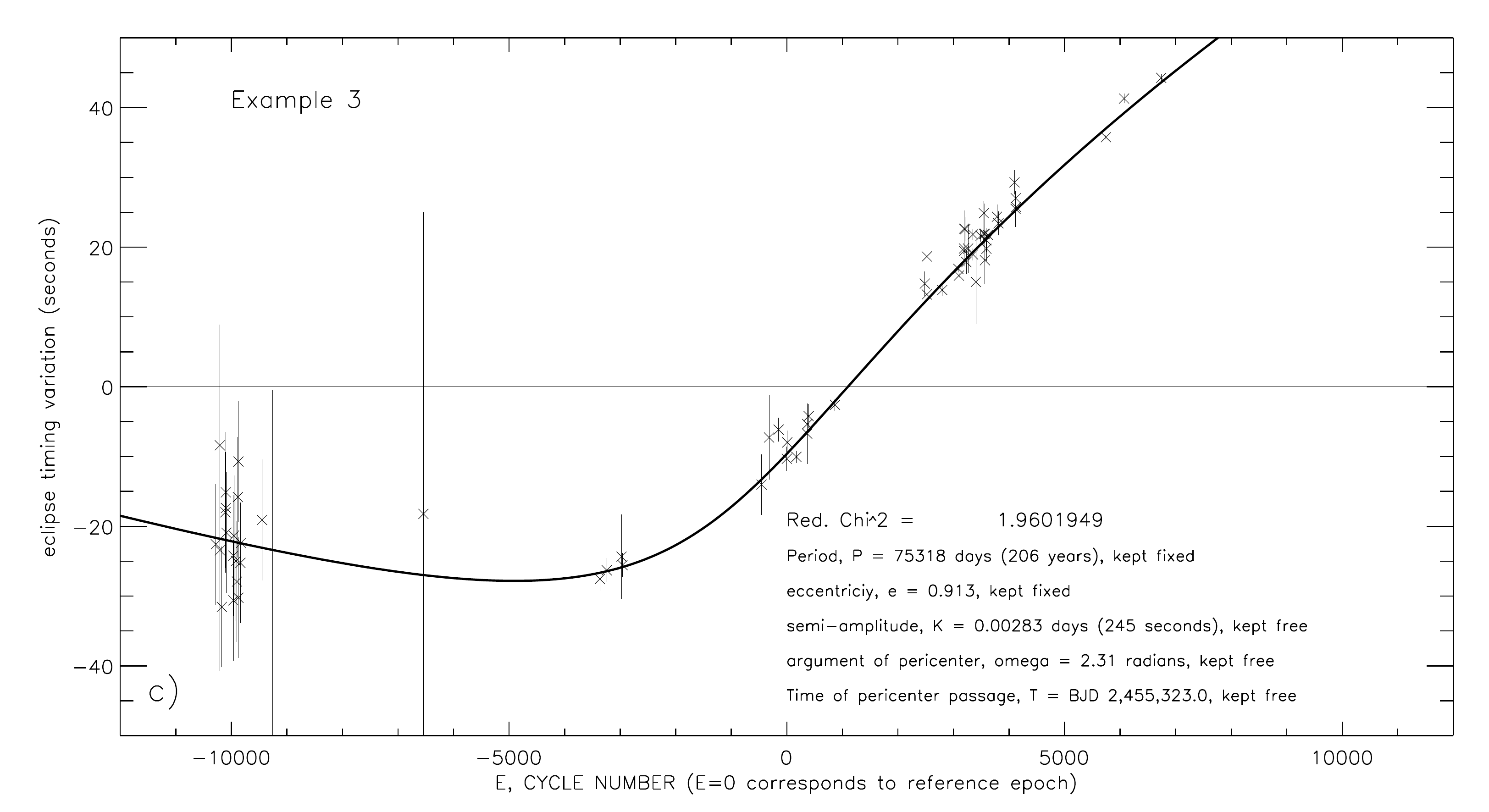}
\includegraphics[angle=0,scale=0.25]{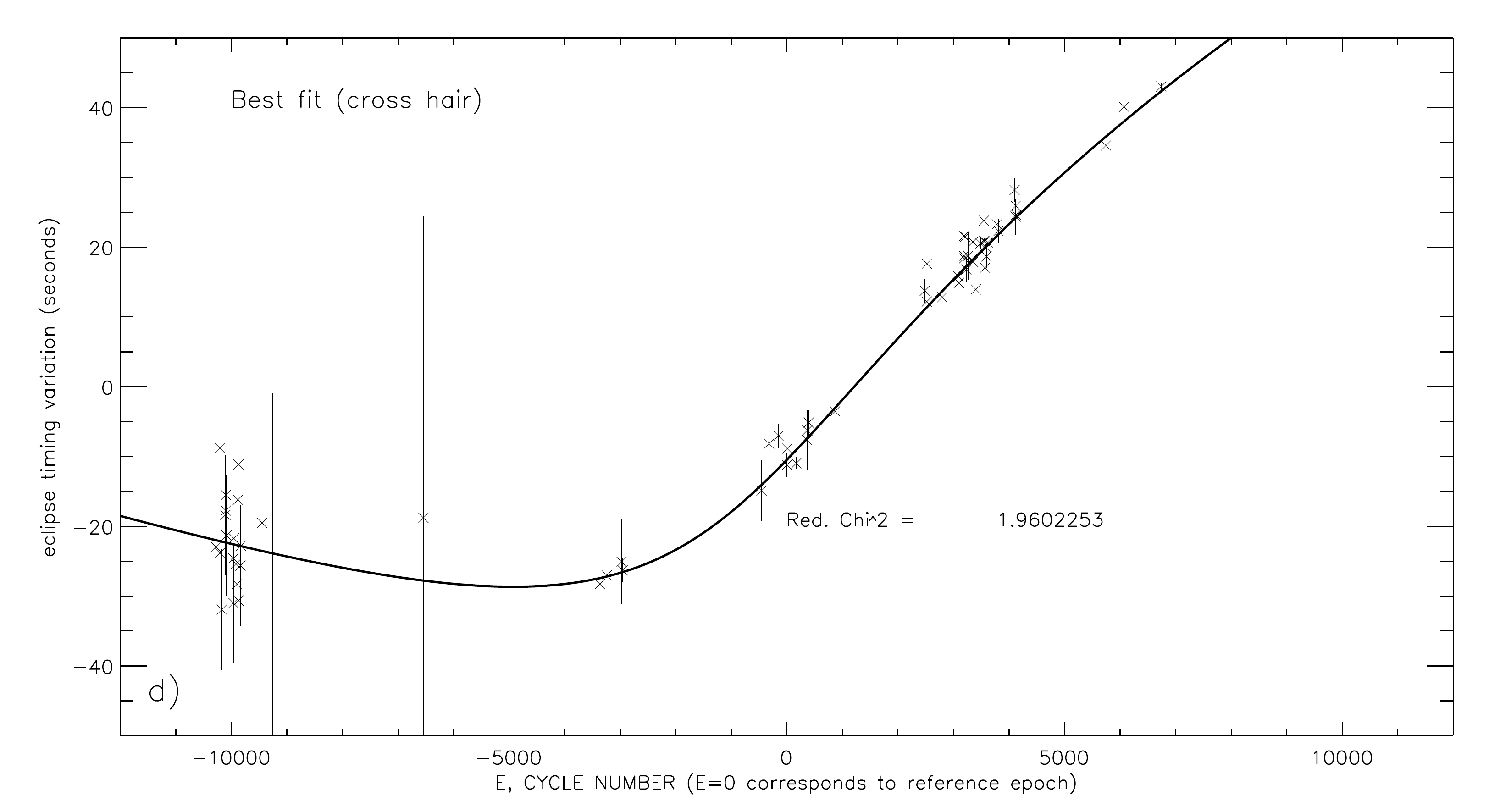}
\caption{Results of considering various models in 
Fig.~\ref{2DScansThreeFigures}. Panels a to c show the models for example 1 to 3. Panel d shows the best-fit model indicated by a cross-hair in Fig.~\ref{2DScansThreeFigures}. All models have a reduced $\chi^2$ statistic within the 1-sigma confidence limit, but the underlying orbital architectures are differing significantly. See text for more details.}
\label{BestFitSolutions4Figs}
\end{figure*}

For this reason, the considered parameters (semi-amplitude, eccentricity and period) span a vast range, making it impossible to place firm confidence intervals on them. From Fig.~\ref{2DScansThreeFigures} possible periods span from $\simeq 2500$ days (6.8 years) to at least 80,000 days (219 years) chosen as our upper cut-off limit in the search procedure. We have tested this result by selecting three significantly different pairs of $(P_{1},e_{1})$ in Fig.~\ref{2DScansThreeFigures}a. We label them as follows: \emph{Example 1}): $(P_{1},e_{1})=(3973~\textnormal{days},0.40)$. \emph{Example 2}): $(P_{1},e_{1})=(15769~\textnormal{days},0.77)$. \emph{Example 3}): $(P_{1},e_{1})=(75318~\textnormal{days},0.91)$. We re-calculated a best-fit model with the $(P_{1}, e_{1})$ parameters held fixed, and remaining 
parameters $(T_{0}, P_{0}, K_{1}, \omega_{1}, T_{1})$ allowed to vary freely (starting from the best-fit solution given by the cross-hair in Fig.~\ref{2DScansThreeFigures}a) to find new optimum values. We show the results of this experiment in Fig.~\ref{BestFitSolutions4Figs}. All models have $\chi^2_{60}$ within the 1-sigma confidence level (1.993) but differ significantly in their orbital periods, eccentricities and semi-amplitudes. Our best-fit model (cross-hair) is shown in Fig.~\ref{BestFitSolutions4Figs}d and Table \ref{bestfitparam_extended}. We calculated the following values for the companion's minimum mass and semi-major axis for our three examples. \emph{Example 1}): $m_1\sin I_1 = 7.6~M_{jup}$, $a_{1}\sin I_1 = 4.0$ au. \emph{Example 2}): $m_1\sin I_1 = 8.5~M_{jup}$, $a_{1}\sin I_1 = 10.0$ au. \emph{Example 3}): $m_1\sin I_1 = 9.7~M_{jup}$, $a_{1}\sin I_1 = 28.4$ au. In light of the large range of possible parameters we omit quoting parameter uncertainties. Minimum mass and semi-major axis for our improved best-fit solution (Fig.~\ref{BestFitSolutions4Figs}d) are given in Table ~\ref{bestfitparam_extended}.

Up to this point our analysis allows us to conclude that the data is not spanning a sufficiently long observing baseline to firmly constrain the parameters of a single companion model. We stress that the model itself could still be valid. With the data currently at hand it is impossible to establish firm confidence intervals on the parameters. Our first solution (comparable with the solution presented in \cite{Almeida2013}) likely represents a local minimum in the $\chi^2_{\nu}$ parameter space, or appears to be a solution within the $1\sigma$ confidence interval characterised by a shallow topology of $\chi^2$ space. In such a case we cannot distinguish isolated models in the continuum of possible solutions. All three panels in Fig.~\ref{2DScansThreeFigures} indicate the existence of local minima with $\chi^2_{\nu}$ statistics close to our first best-fit solution with $\chi^2_{60,0} = 1.98$ (Table~\ref{bestfitparam}). In fact, Fig.~\ref{2DScansThreeFigures} suggests the existence of multiple local minima in the $\chi^2_{\nu}$ space. Since in Fig.~\ref{2DScansThreeFigures} we have not found the 1-sigma confidence level to render as a closed-loop contour line, we suspect that the data can be fit to an infinite number of models each having the \emph{same} statistical significance, but exhibiting significant differences in their orbital architectures. In light of this result any efforts to search for a second companion in Dataset I seems unfruitful.

\subsection{Single companion model - Dataset II}

\begin{figure*}
\includegraphics[angle=0,scale=0.45]{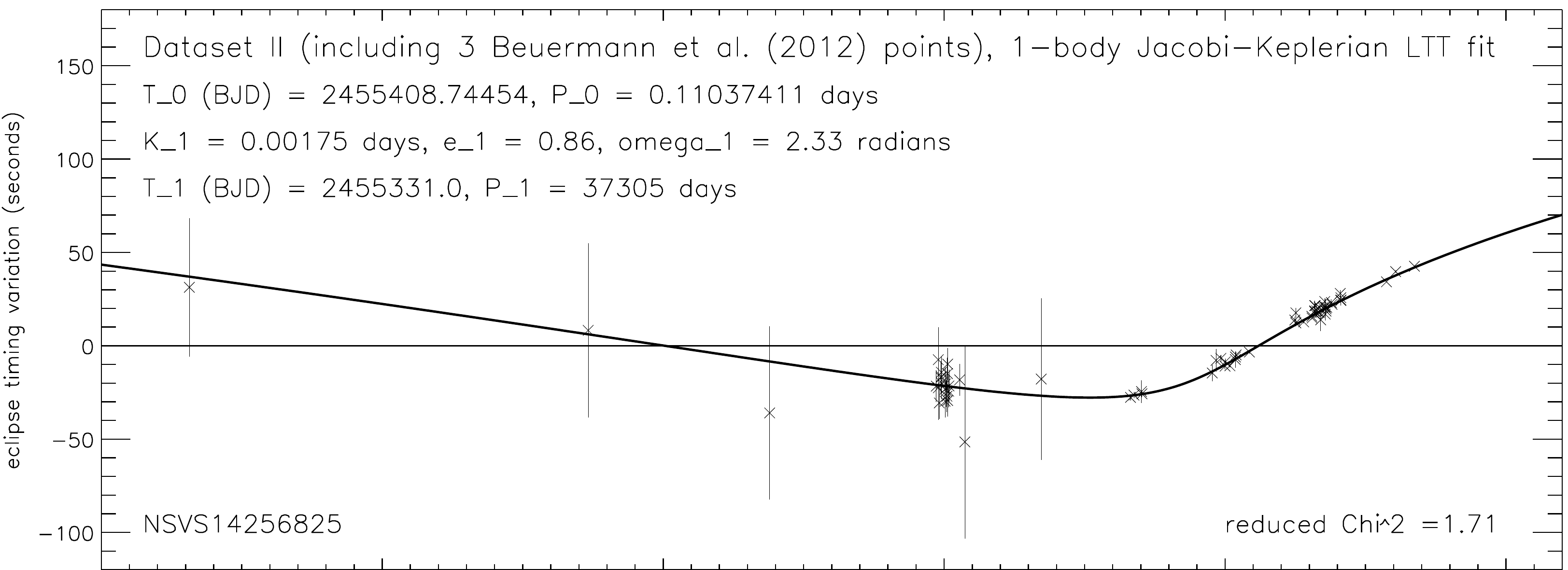}
\includegraphics[angle=0,scale=0.45]{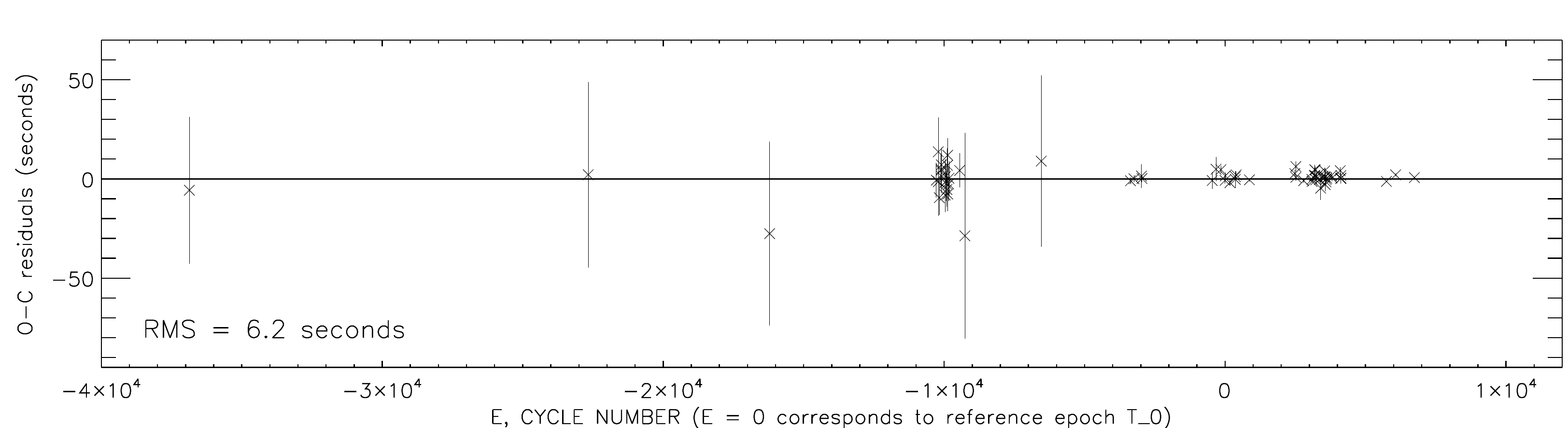}
\caption{Best-fit solution based on Dataset II (see section \ref{dataacquisition}). The $\chi^2_{60,0}$ is smaller due to the inclusion of additional three timing measurements by \cite{Beuermann2012a}. The residual plot does not support the existence of an additional light-travel time signal above the root-mean-square (RMS) of about 6 seconds. Signals with a semi-amplitude smaller than 6 seconds should be treated with caution.} 
\label{BestFitSolutionDataSet2}
\end{figure*}

We have noted that three datapoints from \cite{Beuermann2012a} were not included in the analysis presented in \cite{Almeida2013}. Although they are accurate (placing them well on the linear ephemeris) their timing precision is lower. However, the large timing uncertainty for these points should not disqualify them from being included in the analysis. In principle, the precision of the eclipsing period $P_0$ should increase for a dataset of increased baseline, and could eventually help to constrain any long-period trend. We have repeated our search procedure as outlined previously to find a best-fit model based on dataset II. We show our best-fit solution in Fig.~\ref{BestFitSolutionDataSet2} and state the best-fit parameters within the figure area.

For dataset II, the main characteristics of the Keplerian orbit for the companion are similar to the parameters shown in Table \ref{bestfitparam_extended}. The period, minimum semi-major axis and eccentricity are comparable in both cases. We also explored the topology of $\chi^2$ space for a large region around the best-fit solution and found similar results as discussed previously by generating two-dimensional joint-confidence interval maps. The 1-sigma confidence contour around the best-fit solution extends over a large interval in the period, eccentricity and semi-amplitude.

From examining the residual plot in Fig.~\ref{BestFitSolutionDataSet2} we are not convienced about any additional light-travel time periodicity above the RMS level of about six seconds. A light-travel time signal with amplitude of around six seconds would require a dataset with RMS of about one second or less. Hence, from a qualitative judgment, the data in Dataset II does not currently support the inclusion of five additional parameters corresponding to a second companion. 
The results from examining Dataset II reinforces insufficient coverage of the orbit as presented in \citep{Almeida2013}. Because Dataset II covers two-times the best-fit period found for Dataset I, one would expect Dataset II to constrain the orbital period to a higher degree than for Dataset I. However, this is not the case for the present situation.

\section{Summary and conclusions}

In this work we have carried out a detailed data analysis of timing measurements of the short-period eclipsing binary NSVS14256825. In particular we have examined the one-companion model bearing in mind that additional valid companions should be readily visible in the resulting residuals. On the basis of Dataset I, we first carried out an initial local search for a weighted least-squares best-fit solution. A best-fit model (Table~\ref{bestfitparam}) with $\chi^{2}_{\nu} \simeq 1.98$ resulted in an inner circumbinary companion with orbital characteristics comparable to the short-period companion presented in \cite{Almeida2013}. Extending our search grid of $\chi^2$ parameter space resulted in a similar best-fit $\chi^2_{\nu}$ statistic with significantly different orbital characteristics (Table~\ref{bestfitparam_extended}). We were able to show quantitatively that the present timing data does not allow us to firmly constrain a particular model with well-established parameter confidence limits. In light of this, quoting formal errors for the model parameters seems meaningless. We concluded that the best-fit solution found by \cite{Almeida2013} most likely represents a local minimum. We explain the lack of constraint in the parameters by the limited monitoring baseline over which timing data was acquired. Dataset I represented a baseline of about 5 years. If a periodicity is present, the principle of recurrance should apply, requiring two full orbital periods to be covered in order to establish firm evidence for the presence of a companion. This would correspond to a light-travel time period of at most 2.5 years for Dataset I and 6 years for Dataset II (spanning about 13 years). However, for Dataset I, the data did not allow models with periods shorter than $\simeq 1000$ days. Simultaneously Dataset II does not constrain the period any better than Dataset I.

Our analysis did not allow us to find convincing evidence of a second light-travel time signal. The RMS scatter of timing data around the best-fit model was found to be around 5 seconds. Signals with a semi-amplitude comparable with the measurement uncertainties seem unlikely to be supported by the present data. The claimed second companion in \cite{Almeida2013} has a semi-amplitude of $K_2 \simeq 4.9$ seconds. It is likely that noise was wrongly interpreted as a light-travel time signal. We recommend that a secure detection requires a signal semi-amplitude of at least three times above the noise level (i.e $K \simeq 3 \times \textnormal{RMS}$). Future timing data \citep{Pribulla2012, ParkKMTnet2012} of this system will be important to help constraining the parameters significantly.


\subsubsection*{Acknowledgements}

Research by T.~C.~H is carried out at the Korea Astronomy and Space Science Institute (KASI) under the 2012 KRCF (Korea Research Council for Science and Technology) Young Scientist Research Fellowship Program. K.~G. is supported by Polish NSC, grant N/N203/402739. Numerical computations were partly carried out using the SFI/HEA Irish Centre for High-End Computing (ICHEC) and the PLUTO computing cluster at KASI. Astronomical research at Armagh Observatory is funded by the Department of Culture, Arts and Leisure (DCAL). T.~C.~H and J.~W.~L acknowledges support from KASI grant 2013-9-400-00. J.~H. gratefully acknowledges financial support of the Australian government through ARC Grant DP0774000. R.A.W is supported by a UNSW Vice-Chancellor's Fellowship.

\clearpage

\clearpage

\clearpage

\clearpage

\clearpage

\clearpage

\clearpage

\clearpage

\clearpage

\clearpage

\clearpage

\clearpage

\clearpage

\begin{table*}
\centering
\begin{tabular}{lcccc}
\hline
 & Dataset I & \\
\hline
\hline
 & $\chi^2_{60,0} = 1.98$, $N = 67$, $n = 7$, $\nu = 60$ & \\
\hline
RMS & 5.4 & seconds         \\
\hline
$T_0$         & 2,455,408.74450(36)      &   BJD             \\
$P_0$         & 0.11037415(5)            &   days               \\
\hline
$K_{1}$       &  $0.00023 \pm 0.00005$   &   AU              \\
$e_{1}$       &  $0.3 \pm 0.1$      &   -                \\
$\omega_{1}$  &  $1.7 \pm 0.3$    &   radians             \\
$T_{1}$       &  2,455,197(67)       &   BJD             \\
$P_{1}$       &  $2921 \pm 258$      &   days            \\
\hline

$m_{1}\sin I_{1}$ & $6.7 \pm 0.9$   &  $M_{jup}$       \\
$a_{1}\sin I_{1}$ & $3.3 \pm 0.6$         &  AU                       \\
$e_{1}$           & $0.3 \pm 0.1$   & - \\
$\omega_{1}$      & $(1.7 + \pi) \pm 0.3$    & radians \\
$P_{1}$           & $2921 \pm 258$   & days \\
\hline
\end{tabular}
\caption{Best-fit parameters for the one-companion LTT model of NSVS14256825
corresponding to Fig.~\ref{1bodyLTTJacBestFit}. RMS measures the root-mean-square scatter of the data around the best fit model. We quote 
formal uncertainties obtained from the covariance matrix provided by \texttt{MPFIT}. Uncertainties for the minimum mass and semi-major axis 
(relative to the binary mass center) of the companion were derived via error propagation \citep{BevingtonRobinson2003}.}
\label{bestfitparam}
\end{table*}

\clearpage

\begin{table*}
\centering
\begin{tabular}{lcccc}
\hline
& Dataset I & \\
\hline
\hline
 & $\chi_{60,0}^2 = 1.96$, $N = 67$, $n = 7$, $\nu=60$ & \\
\hline
RMS & 5.3 & seconds         \\
\hline
$T_0$         & 2,455,408.74455(41)      &   BJD             \\
$P_0$         & 0.11037411(6)            &   days               \\
\hline
$K_{1}$       &  $0.00169$   &   AU              \\
$e_{1}$       &  $0.85$      &   -                \\
$\omega_{1}$  &  $2.33$    &   radians             \\
$T_{1}$       &  2,455,330       &   BJD             \\
$P_{1}$       &  $34263$      &   days            \\
\hline

$m_{1}\sin I_{1}$ & $9.8$   &  $M_{jup}$       \\
$a_{1}\sin I_{1}$ & $16.8$         &  AU                       \\
$e_{1}$           & $0.85$   & - \\
$\omega_{1}$      & $(2.33 + \pi)$    & radians \\
$P_{1}$           & $34263$   & days \\
\hline
\end{tabular}
\caption{Similar to Table \ref{bestfitparam}, but this time the best-fit (Fig.~\ref{BestFitSolutions4Figs}d) is obtained from the extended search case by randomly generate initial guesses from a region spanning a larger interval of the parameters (mainly $K_{1}, e_{1}$ and $P_{1}$). In Fig.~\ref{2DScansThreeFigures} we show the best-fit parameters for $P_{1}, e_{1}$ and $K_{1}$ as a cross-hair. Formal parameter uncertainties are omitted (see text for details).}
\label{bestfitparam_extended}
\end{table*}

\end{document}